\documentclass[aps,prd,twocolumn,floatfix,nofootinbib,superscriptaddress,preprintnumbers]{revtex4-2}
\usepackage{amsmath,amssymb,mathtools}
\usepackage{graphicx,siunitx,physics,booktabs,multirow}
\usepackage{color}
\usepackage[bookmarks, linktocpage, colorlinks = true, linkcolor = blue, urlcolor  = blue, citecolor = blue, anchorcolor = green, hyperindex = true, hyperfigures]{hyperref}

\newcommand{\density}{n_{\mathrm{B}}}
\newcommand{\chem}{\mu_{\mathrm{B}}}
\newcommand{\energy}{\varepsilon}

\begin{document}
\title{Signature of hadron-quark crossover in binary-neutron-star mergers}
\author{Yuki Fujimoto}
\affiliation{Institute for Nuclear Theory, University of Washington, Box 351550, Seattle, WA 98195, USA}
\author{Kenji Fukushima}
\affiliation{Department of Physics, The University of Tokyo, 7-3-1 Hongo, Bunkyo-ku, Tokyo 113-0033, Japan}
\author{Kenta Hotokezaka}
\affiliation{Research Center for the Early Universe (RESCEU), Graduate School of Science, The University of Tokyo, Tokyo 113-0033, Japan}
\author{Koutarou Kyutoku}
\affiliation{Department of Physics, Graduate School of Science, Chiba University, Chiba 263-8522, Japan}
\affiliation{Interdisciplinary Theoretical and Mathematical Sciences Program (iTHEMS), RIKEN, Wako, Saitama 351-0198, Japan}

\date{\today}

\preprint{INT-PUB-24-041}

\begin{abstract}
 We study observational signatures of the hadron-quark crossover in binary-neutron-star mergers by numerical-relativity simulations with various mass configurations. We employ two equations of state (EoSs) for matter consistent with inference from the observational data. In the crossover scenario the EoS is softened in a density realized in binary-neutron-star mergers and is smoothly continued to quark matter. In the phase transition scenario without crossover, the EoS remains stiff and a first-order phase transition takes place in a density out of reach of mergers. A GW170817-like system forms a remnant massive neutron star in both scenarios, and it collapses into a black hole only in the crossover scenario due to the softening while gravitational-wave emission is strong. This difference is clearly reflected in the sudden shutdown of gravitational waves. For a given EoS, the lifetime of the merger remnant is determined primarily by the total mass of the system. Identifying these features in a variety of future events with the next generation of ground-based gravitational-wave detectors will enable us to clarify details of hadron-quark transition. The mass of the accretion disk surrounding the remnant black hole is affected not only by the lifetime of the remnant but also by the mass ratio of the system. Electromagnetic emission associated with the disk outflow will also be useful for detailed investigation of the hadron-quark transition.
\end{abstract}

\maketitle

\section{Introduction}

The nature of supranuclear-density matter is one intriguing and yet unresolved problem in quantum chromodynamics (QCD). Arguably, in this context, the most important question to be addressed is the clarification of the QCD phase structure on the plane with the temperature $T$ and the baryon density $\density$ (or the baryon chemical potential $\chem$). On the one hand, hadrons are appropriate degrees of freedom to describe the low-$T$ and low-$\chem$ dynamics. Quarks must take over the physical relevance, on the other hand, at the high-$T$ and/or high-$\chem$ regime. The question is how these two distinct realizations of degrees of freedom are interchanged. For $\chem\ll T$, the scaling analysis in the lattice-QCD simulation has revealed that hadrons and quarks are not separated by a sharp phase transition (PT) but connected by a smooth crossover (CO)~\cite{Aoki:2006we}. However, the lattice-QCD simulation cannot access the low-$T$ and high-$\chem$ regions due to the notorious sign problem; see Ref.~\cite{Gattringer:2016kco} for a review on some selected approaches to the sign problem.

Since 2017, gravitational waves from binary neutron stars have widely been recognized as a useful tool to constrain the properties of supranuclear-density matter based on astronomical observations. The inspiral gravitational waveforms observed in GW170817~\cite{LIGOScientific:2017vwq} delivered information about the tidal deformability of premerger neutron stars~\cite{LIGOScientific:2018cki,De:2018uhw,Narikawa:2019xng}. This event constrained the equation of state (EoS) at $\density\sim 2n_\text{sat}$ where $n_\text{sat}\simeq\SI{0.16}{\per\cubic\femto\meter}$ represents the nuclear saturation density; the EoS in this density region characterizes neutron stars with the canonical mass of $1.3$--$1.4M_\odot$. Although information from later events such as GW190425~\cite{LIGOScientific:2020aai} was not as useful as that from GW170817, it is highly promising that future observations with improved detectors will significantly tighten the constraints on the dense matter properties around $\sim 2n_\text{sat}$ (see, e.g., Refs.~\cite{HernandezVivanco:2019vvk,Wysocki:2020myz,Landry:2020vaw,Gupta:2022qgg,Walker:2024loo}).

It is one of the future goals in gravitational-wave astronomy to infer stronger constraints on ultrahigh-density matter properties at $\density > 2n_\text{sat}$. For this purpose, it is desirable to observe postmerger remnant massive neutron stars with increased density (see, e.g., Refs.~\cite{Bauswein:2011tp,Hotokezaka:2013iia,Takami:2014zpa,Bernuzzi:2015rla}). However, because the frequency of gravitational waves from the postmerger remnant is as high as a few \si{\kilo\hertz} reflecting its dynamical timescale, the postmerger gravitational-wave signals remain elusive even for GW170817 at \SI{40}{Mpc} distance~\cite{LIGOScientific:2017fdd}. The situation should be improved once third-generation \cite{Maggiore:2019uih,Gupta:2023lga} and/or specially-designed detectors \cite{Martynov:2019gvu,Ackley:2020atn,Srivastava:2022slt} become available. Having this future direction in mind, researchers in various fields have investigated the prospects for determining the nature of ultrahigh-density matter from postmerger gravitational waves, particularly focusing on the scenario with strong first-order PT (see, e.g., Refs.~\cite{Radice:2016rys,Most:2018eaw,Bauswein:2018bma,Weih:2019xvw,Liebling:2020dhf} and also the 1PT part in Ref.~\cite{Hensh:2024onv}). These studies found that sufficiently strong first-order PT may be identified via comparisons of pre- and postmerger signals such as unexpectedly high peak frequency of the postmerger emission.

Recent studies of nuclear-matter properties rather favor an alternative scenario without strong first-order PT at moderate density relevant to astronomical phenomena~\cite{Annala:2019puf,Fujimoto:2019hxv,Fujimoto:2021zas,Fujimoto:2024cyv}. Such an EoS construction with smooth interpolation between nuclear and quark matter is commonly referred to as the hadron-quark CO scenario which was theorized in Ref.~\cite{Masuda:2012kf}, whereas the first-order PT is still consistent with phenomenology~\cite{Benic:2014jia, Christian:2023hez, Li:2024sft}. For the moment, the \emph{ab initio} approaches are possible for hadronic matter at low density and quark matter at high density; namely, chiral effective field theory ($\chi$EFT) and perturbative QCD (pQCD), respectively, at sufficiently low and high densities. Although it is impossibly difficult to derive the properties of intermediate-density matter from first principles, various astronomical observations and terrestrial nuclear experiments allow us to impose meaningful constraints. The uncertainty range in the EoS is getting narrower, and we believe that the acceptable strength of the first-order PT in the intermediate density range should be severely limited in the near future, unless the first-order PT might be evidenced. Moreover, the existence of neutron stars with the mass $\approx 2M_\odot$ \cite{Antoniadis:2013pzd,NANOGrav:2017wvv,Fonseca:2021wxt}\footnote{There also exists a claimed observation of more massive pulsars with relatively large error margins~\cite{Romani:2022jhd}.} as well as the small tidal deformability from GW170817~\cite{LIGOScientific:2017vwq, LIGOScientific:2018cki} requires rapid stiffening of intermediate-density matter. This observation also suppresses parameter space of first-order PT, though it cannot be completely ruled out~\cite{Alford:2013aca, Christian:2023hez, Li:2024sft}.

Without the PT, the smooth onset of strongly-interacting quarks with repulsive vector interaction could naturally explain the stiffening at intermediate density; see a review~\cite{Baym:2017whm}. This non-perturbative nature of quark matter could be interpreted as a dual of strongly-interacting nuclear matter, so that a quarkyonic model can give an account for the stiffening~\cite{McLerran:2007qj,McLerran:2018hbz}; see also a solvable quarkyonic model in Ref.~\cite{Fujimoto:2023mzy}. Moreover, we point out that the diquark condensates in color superconductors could fill in the gap between nuclear and quark matter, as first speculated based on global symmetries in Ref.~\cite{Schafer:1998ef}. Actually, the diquark condensates are of practical importance for the EoS matching in the CO scenario~\cite{Baym:2017whm}, which may also connect the non-strange sector~\cite{Fukushima:2015bda,Fujimoto:2019sxg}. For a recent attempt to extract the diquark condensates from the neutron star observation, see Ref.~\cite{Kurkela:2024xfh}. Although the CO is a natural candidate, a possibility of topological first-order PT is under intensive debates~\cite{Alford:2018mqj,Cherman:2018jir,Hirono:2018fjr,Hayashi:2023sas}. Thus, in view of those theoretical controversies, it is becoming more and more important to understand the influence of hadron-quark CO on the dynamics of binary-neutron-star mergers and observable signals.

In this paper, we investigate the effects of hadron-quark PT/CO on gravitational waves by performing numerical-relativity simulations of binary-neutron-star mergers systematic in terms of mass configurations. For this purpose, we employ two representative EoSs for matter with and without softening associated with CO toward quark matter. To make our models realistic, our EoSs are chosen to be consistent with the results derived by $\chi$EFT in the low-density side and pQCD in the high-density side.
We have reported in the preceding Letter~\cite{Fujimoto:2022xhv} that the merger remnant collapses into a black hole earlier in the case of CO than that of strong first-order PT at very high density.  Observations of neutron stars suggest a further stiffening of the EoS at densities realized in the core of neutron stars disfavoring a small velocity of sound ~\cite{Brandes:2023hma, Komoltsev:2024lcr}, while the CO scenario is more flexibly compatible.
Thus, if the gravitational collapse is inferred from the gravitational-wave observation, the CO scenario will be favored modulo understanding of the EoS for matter representative of premerger neutron stars. Because we previously made the claim based only on a limited number of binary models with the total mass of $2.75M_\odot$ motivated by GW170817~\cite{LIGOScientific:2017vwq}, we extend our preceding studies by simulating binary models with a variety of masses of the components. As we will explain in the present paper, this extension reveals that the compilation of various binary neutron star systems with different total masses from future observations can delineate the transition scenario in detail.

This paper is organized as follows. We first explain our EoS constructions according to two representative scenarios in Sec.~\ref{sec:eos}. We then compare our constructed EoS with the Bayesian inference in Sec.~\ref{sec:bayes}. Section~\ref{sec:sim} presents the results of our merger simulations and their implications. Discussions about systematic uncertainties follow in Sec.~\ref{sec:sys} referring to the results of additional simulations. Finally, Sec.~\ref{sec:summary} is devoted to a summary. Throughout this paper, the speed of light $c$ is set to unity.

\section{Construction of the equation of state at zero temperature} \label{sec:eos}

We employ the standard piecewise polytropes~\cite{Read:2008iy} as our fiducial model of zero-$T$ EoS in our numerical simulations. The pressure $P_\mathrm{cold}$ at $T=0$ is parametrized as a series of power-law functions of the rest-mass density $\rho$ in the following form:
\begin{equation}
    P_\mathrm{cold} (\rho) = K_i \rho^{\Gamma_i}
\end{equation}
for the $i$-th density interval of $\rho_{i-1} \le \rho < \rho_i$. Note the relation $\rho = m_\mathrm{B} \density$, where the baryon mass $m_\mathrm{B}$ is approximated by the atomic mass unit $m_\mathrm{u}$. Once we specify the values of $K_1$, $\rho_i$, and $\Gamma_i$ (see below for concrete values), the remaining $K_i$'s are determined uniquely by the requirement of the pressure continuity. The energy density $\varepsilon_\mathrm{cold}$ for the same interval is derived by the first law of thermodynamics as
\begin{equation}
    \varepsilon_\mathrm{cold} (\rho) = (1 + a_i) \rho + \frac{K_i}{\Gamma_i - 1} \rho^{\Gamma_i}\,,
\end{equation}
where $a_i$ is an integration constant determined also by the continuity condition. The baryon chemical potential is given by the relation $\chem = (\varepsilon_\mathrm{cold} + P_\mathrm{cold})/\density$, and this is related to the specific enthalpy $h_\mathrm{cold}$ by $\chem = m_\mathrm{B} h_\mathrm{cold}$. Below, the subscript ``cold'' is suppressed unless otherwise noted.

In this work, we contrast the two scenarios: the EoS with CO and with the strong first-order PT\@. For the scenario with the first-order PT, in practice, it is indistinguishable from the pure hadronic EoS since the density of the transition is higher than the density that can be reachable inside neutron stars. In other words, the strong first-order PT can be excluded at low densities if the EoS is not rapidly stiffened immediately above the saturation density.

\begin{table*}
    \caption{Parameters of the piecewise polytropes. We set $K_1 = \num{3.99873692e-8} (\si{\gram\per\cubic\cm})^{1-\Gamma_1}$, $\rho_0 = 0$, and $\rho_4 \to \infty$ (the first-order PT point is out of the reachable density). The radius of a neutron star is \SI{12.1}{\km} for $1.3$--$1.7M_\odot$ and marginally close to \SI{12.0}{\km} for $1.25M_\odot$ irrespective of the transition scenarios (see Table \ref{tab:model} for models considered in this study). The maximum mass of a cold, spherical neutron star is $1.98M_\odot$ for CO and $2.31M_\odot$ for PT.} \label{tab:pwp}
    \centering
    \begin{tabular}{c|ccccccc}
    \toprule
    Model & $\Gamma_1$ & $\rho_1 (\si{\gram\per\cubic\cm})$ & $\Gamma_2$ & $\rho_2 (\si{\gram\per\cubic\cm})$ & $\Gamma_3$ & $\rho_3 (\si{\gram\per\cubic\cm})$ & $\Gamma_4$ \\
    \midrule
    Crossover (CO) & \multirow{2}{*}{\num{1.3562395}} & \multirow{2}{*}{\num{1.022e14}} & \multirow{2}{*}{\num{2.64258}} & \multirow{2}{*}{\num{4.29e14}} & \multirow{2}{*}{\num{3.5}} & \multirow{2}{*}{\num{8.56e14}} & \num{1.45303} \\
    First-order phase transition (PT) & & & & & & & \num{2.9} \\
    \bottomrule
    \end{tabular}
\end{table*}

\begin{figure}
    \includegraphics[width=0.95\linewidth]{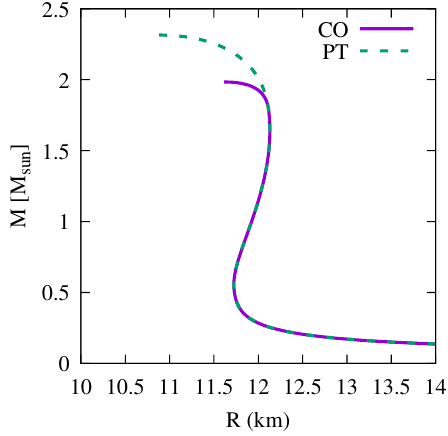}
    \caption{Mass-radius relation for the CO and PT scenarios. Both curves are terminated at the maximum-mass configurations.}
    \label{fig:mr}
\end{figure}

We use the same EoSs as those constructed in the preceding \textit{Letter}. Parameters of our EoSs are summarized in Table~\ref{tab:pwp}, and the corresponding mass-radius relations are shown in Fig.~\ref{fig:mr}. Let us briefly review the procedure we adopted previously below.
\begin{itemize}
  \item{$i=1$:} The first interval represents overall behavior at subsaturation density with $\Gamma_1 = \num{1.3562395}$ and $K_1 = \num{3.99873692e-8} (\si{\gram\per\cubic\cm})^{1-\Gamma_1}$ \cite{Read:2008iy}. We also set $\rho_0 = 0$ and $a_1 = 0$.
  \item{$i=2$:} The next interval is chosen to satisfy the constraint imposed by $\chi$EFT calculations up to $\energy \sim \SI{0.25}{GeV.fm^{-3}}$. The fit results in $\Gamma_2=\num{2.64258}$.
  \item{$i=3$:} The third interval is manually constructed to satisfy various astronomical and terrestrial constraints. We specifically choose $\Gamma_3=\num{3.5}$.
  \item{$i=4$:} We fix the highest-density EoS from the pQCD calculations down to $\energy \sim \SI{1}{GeV.fm^{-3}}$ by taking a relatively large value for the renormalization scale, i.e., $\bar{\Lambda} = 4\mu$ (that is a characteristic scale in the strong coupling constant set by the quark chemical potential $\mu=\chem/3$), leading to $\Gamma_4=\num{1.45303}.$
\end{itemize}
The steps for $i=3$ and $i=4$ need further explanations. For $i=3$, we anchor the EoS to the \textit{ab initio} calculations at low and high densities, so that the likely EoS candidate should interpolate them. This part has large uncertainties but as explained in the introduction, inspiral gravitational waveforms in the future observations should impose stronger constraints. Therefore, it is a sensible and pragmatic strategy to take a single polytrope with a certain value of $\Gamma_3$ for the moment, instead of scanning over various EoS candidates, and refine the simulations after we have more data at our disposal in the near future. Previously, we chose $\Gamma_3=3.5$ and we keep using the same value in this work. It should be noted that this choice is motivated by the analysis of the neutron star observations in the neural-network method~\cite{Fujimoto:2019hxv,Fujimoto:2021zas,Fujimoto:2024cyv} which concludes a rather soft EoS with $\Gamma\simeq 3.5$. To identify the EoS for $i=3$ uniquely, we must specify not only $\rho_2$ but also $\rho_3$.

\subsection{Crossover Scenario --- CO}

In the piecewise polytropic parametrization, the EoS construction is conceptually straightforward especially for the CO scenario. Once we fix the highest-density part, i.e., the $i=4$ interval in this case, $\rho_3$ is uniquely determined by the crossing point between two EoSs before and after the CO\@. Now, we must choose $\Gamma_4$ for the highest-density interval, and we do so to match the prediction of pQCD for the CO scenario, as stated above. For our choice of $\Gamma_4$, we find $\rho_3=\SI{8.56e14}{\gram\per\cubic\cm}$ as listed in Table~\ref{tab:pwp}.

\begin{figure}
    \centering
    \includegraphics[width=.99\columnwidth]{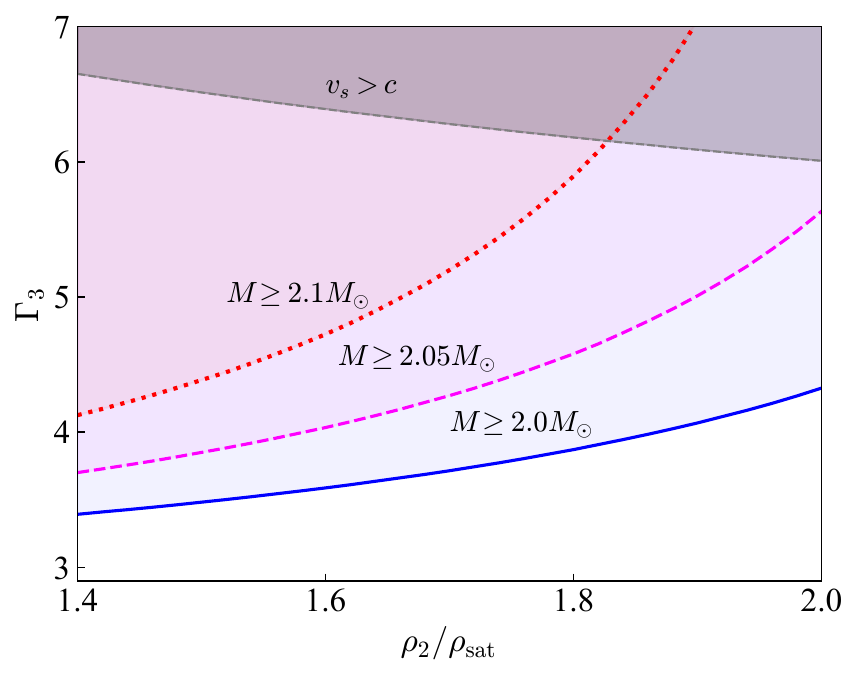}
    \caption{Parameters and the maximum-mass boundaries in the crossover scenario. The larger value of the maximum mass requires larger $\Gamma_3$ but too large $\Gamma_3$ is excluded by the causality bound.}
    \label{fig:param_CO}
\end{figure}

The remaining parameter is $\rho_2$, and the allowed region of $\rho_2$ is correlated with the choice of $\Gamma_3$. For a given set of $\rho_2$ and $\Gamma_3$, the CO EoS is derived, from which we can compute the maximum mass of the neutron star as shown in Fig.~\ref{fig:param_CO}. Our choice of $\Gamma_3=3.5$ corresponds to a soft EoS, and the high-density EoS in the pQCD branch is also soft. Therefore, we should take $\rho_2 \lesssim 1.6\rho_{\mathrm{sat}}$ to guarantee that the maximum mass of the neutron star reaches twice the solar mass, where $\rho_{\mathrm{sat}}$ is the rest-mass density at nuclear saturation. Physically, $\rho_2$ is interpreted as a validity limit of the $\chi$EFT results. In this work, we choose the nearly boundary value, $\rho_2=1.6\rho_{\mathrm{sat}}$, which corresponds to $M_\mathrm{max} = 1.98M_\odot$.
Although this may appear to be smaller than the currently observed $M_{\max}$~\cite{Fonseca:2021wxt}, our $M_{\max}$ is still consistent within the 95\% credible interval.
As stated before, if the error bar in the maximum-mass determination is reduced in the future observation, these parameters should be readjusted.

\subsection{First-order Phase Transition Scenario --- PT}

\begin{figure}
    \centering
    \includegraphics[width=.95\columnwidth]{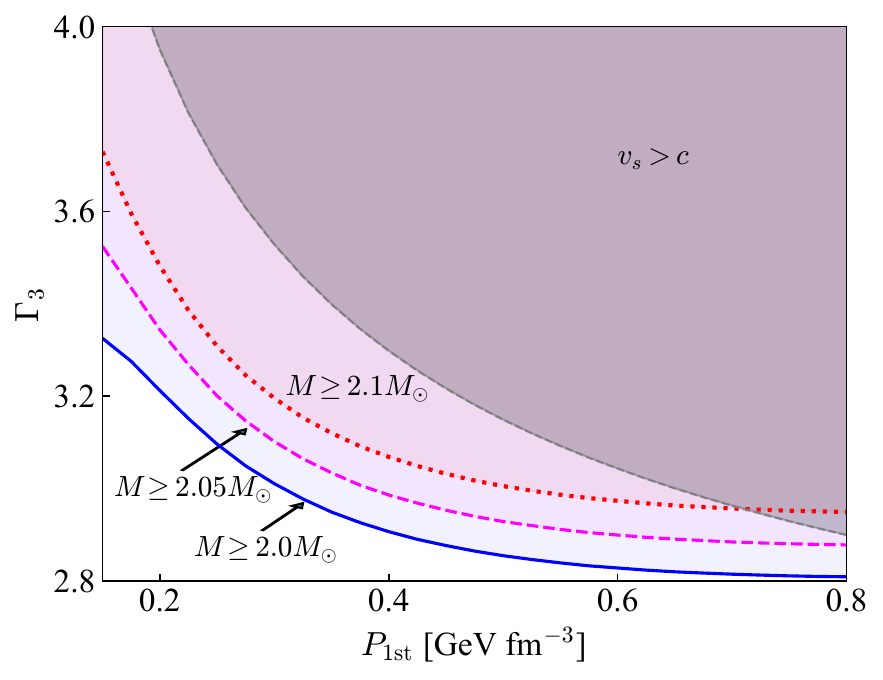}
    \caption{Parameters and the maximum-mass boundaries in the first-order phase transition scenario.}
    \label{fig:param_1st}
\end{figure}

For the PT scenario, in which the strong first-order PT is assumed to occur, we need one more parameter $P_\text{1st}$ for the location of the PT\@. For simplicity, let us choose the same value of $\rho_2$ as in the CO scenario. We assume the Maxwell construction of the first-order PT and the EoS in the energy-pressure plane has a plateau at $P=P_\text{1st}$ attached to the pQCD branch in the high-density side. Figure~\ref{fig:param_1st} shows the allowed parameter region for respective values of the maximum mass. Obviously, the maximum-mass boundaries are insensitive to $P_\text{1st}$ if it is high enough. This is simply because the PT takes place at very high density beyond the reach of stable neutron stars. However, in this case, if the EoS is extended to such high density with the fixed value of $\Gamma_3=3.5$, the causality is violated soon. To avoid the problem of the sound speed exceeding the speed of light, we slightly bend the EoS; that is, we decrease the index from $\Gamma_3=3.5$ to $\Gamma_4=\num{2.9}$ for $\rho>\rho_3$.

Now, the problem is how to make an educated guess for $P_\text{1st}$. From Fig.~\ref{fig:param_1st}, we could safely take
a sufficiently low value of $P_\text{1st}$ for
$\Gamma_3\sim 3.5$ to respect the causality and satisfy $M\ge 2M_\odot$. However, if $P_{\mathrm{1st}}$ is too low, the jump to the pQCD branch is inevitably suppressed, and only a weak first-order PT can be realized.
Therefore, for the purpose of clear numerical comparison, we shall contrast the EoS with CO and the EoS with the strong first-order PT located around $P_\text{1st} = \SI{0.5}{GeV.fm^{-3}}$ by decreasing $\Gamma_3$ from $3.5$ to $2.9$.
It should be noted that, in practice, the corresponding energy density even for a weak first-order PT around $P_\text{1st}\approx\SI{0.2}{GeV.fm^{-3}}$ is too high for the system in the gravitational wave simulation to reach it.  In other words, according to our EoS construction, the PT EoS can just be regarded as an extension of the conventional hadronic EoS without softening in the density region relevant to the binary-neutron-star merger.

\section{Comparison with Bayesian inference}
\label{sec:bayes}

We will gain a deeper insight in the EoS construction by comparing the results from Bayesian inference with the CO and PT EoSs discussed in the previous section. The construction above was based on strong assumptions on the pQCD results. Namely, we fixed the renormalization scale in the pQCD results with a specific value of coefficient, $\bar{\Lambda}/\mu = 4$ (where $\bar{\Lambda}$ is a scale parameter in the running coupling and $\mu$ is the quark chemical potential), and also extended the applicable limit of the pQCD down to a relatively low value $\energy \sim \SI{1}{GeV.fm^{-3}}$. These two assumptions are intertwined with each other because the running coupling constant in QCD becomes smaller for the larger value of $\bar{\Lambda}/\mu$, so that the pQCD can be valid down to the energy density as low as $\energy \sim \SI{1}{GeV.fm^{-3}}$ for $\bar{\Lambda}/\mu=4$, though it is not the case for $\bar{\Lambda}/\mu=1$.

As is well known in this context (see Ref.~\cite{Blaizot:2003iq} for the similar convergence problem and the possible remedy in hot QCD), there is an ambiguity in the choice of $\bar{\Lambda}$ arising from the truncation of the perturbative series at a finite order. The uncertainty in $\bar{\Lambda}$ is customarily estimated by varying it within $\mu \leq \bar{\Lambda} \leq 4\mu$ around its typical value, $\bar{\Lambda} = 2\mu$. So, the above construction corresponds to the upper bound on the customary range of $\bar{\Lambda}$. In this section, we take the conservative values for the pQCD branch and incorporate them in the Bayesian inference. The aim of these analyses is twofold: The first is to show that the conservative values for the pQCD will also lead to the effect on the EoS that is consistent with the description in the previous section. Second, it would be instructive to see where in the currently allowed bounds of the EoS the constructed EoS falls into.

\subsection{Methods}

We perform Bayesian inference, in which the probability distribution for the unknown parameters $\theta$ is obtained as a posterior probability $\Pr(\theta|D)$ for given data $D$ from Bayes's theorem:
\begin{equation}
    \Pr(\theta | D) = \frac{\Pr(D | \theta) \Pr (\theta)}{\Pr (D)}\,.
\end{equation}
Here, $\theta$ and $D$ collectively refer to the undetermined parameters in the theory and the given measurement data, respectively; in the present context, $\theta$ is the parameters of the EoS, and $D$ is the observables of neutron stars. The prior $\Pr(\theta)$ is modeled by the parameters as explained below. For the evaluation of the likelihood $\Pr(D|\theta)$, we follow the constant-likelihood approach developed in Refs.~\cite{Altiparmak:2022bke, Jiang:2022tps}, which can be regarded as a Bayesian extension of Ref.~\cite{Annala:2019puf}. In the constant-likelihood approach, the likelihood is implemented as a step function to impose a cutoff on variables so that the Bayesian update simply rejects models that do not fulfill the condition without performing the numerically demanding Monte Carlo integration. See also Ref.~\cite{Fujimoto:2024pcd}.

\subsubsection{Prior belief}

For the prior distribution, we randomly sample EoSs by modeling them with the piecewise sound-speed parametrization introduced in Ref.~\cite{Annala:2019puf}. We use the standard crust EoS at low $\chem$~\cite{Baym:1971pw}, and use the EoS from the $\chi$EFT~\cite{Drischler:2020fvz} up to $\chem = \mu_{\chi\mathrm{EFT}}$, where we choose $\mu_{\chi\mathrm{EFT}} = \SI{1.0}{GeV}$. The density, pressure, and sound velocity at $\mu_{\chi\mathrm{EFT}}$ are $n_{\chi\mathrm{EFT}} \simeq \SI{0.24}{fm^{-3}}$, $P_{\chi\mathrm{EFT}}\simeq 8.8_{-1.5}^{+1.4}\times\SI{e-3}{GeV.fm^{-3}}$, and $v^2_{s, \chi\mathrm{EFT}} = 0.10\pm0.02$, respectively. The error corresponds to $\pm 1 \sigma$ credible region presented in Ref.~\cite{Drischler:2020fvz}, and we sample the equal number of EoSs at average and at $\pm 1 \sigma$ credible interval (CI) in the prior distribution. We choose the number of the segments as $N=5$ and then randomly sample parameters $\mu_i$ and $v_{s,i}^2$ ($i=1,\ldots,N-1$) from the uniform distributions $U(\mu_{\chi\mathrm{EFT}}, \mu_{\rm pQCD})$ and $U(0, 1)$, respectively, where we choose $\mu_{\rm pQCD} = \SI{2.6}{GeV}$. Then, we linearly interpolate between adjacent segment boundaries at $(\mu_i, v_{s,i}^2)$ with $\mu_0 = \mu_{\chi\mathrm{EFT}}$ and $\mu_N = \mu_{\rm pQCD}$:
\begin{equation}
    v_s^2(\chem) = \frac{(\mu_{i+1} - \chem) v_{s,i}^2 + (\chem - \mu_i) v_{s,i+1}^2 }{\mu_{i+1} - \mu_i}
\end{equation}
for $\mu_i \le \chem < \mu_{i+1}$. Using this linearly-interpolated $v_s^2(\chem)$, the complete thermodynamic function $P(\chem)$ can be constructed through the following integration:
\begin{equation}
  \begin{split}
    &P(\chem) = P_{\chi\mathrm{EFT}} \\
    &\quad + n_{\chi\mathrm{EFT}} \int_{\mu_{\chi\mathrm{EFT}}}^{\chem} \!\! d\mu' \exp\left[\int_{\mu_{\chi\mathrm{EFT}}}^{\mu'} \!\!d\mu''\frac{1}{\mu'' v_s^2(\mu'')}\right]\,.
  \end{split}
  \label{eq:Pmu}
\end{equation}
This expression can be understood from
\begin{equation}
    v_s^2(\chem) = \frac{\partial P}{\partial\chem} \biggl(\frac{\partial \varepsilon}{\partial\chem}\biggr)^{-1} = \frac{\density}{\chem (\partial\density/\partial\chem)}\,,
\end{equation}
which leads to $1/(\mu v_s^2)=\partial\ln \density/\partial\chem$. Then, it is easy to see that the integrand in the $\mu'$-integration is $\density(\mu')/n_{\chi\mathrm{EFT}}$.

\subsubsection{Incorporating observational constraints by a constant likelihood}

For the Bayesian update step, we use the information of measurements of massive pulsars, mass-radius $M$-$R$ of pulsars J0030+0451 and J0740+6620 from NICER, and the binary tidal deformability $\tilde{\Lambda}$ from the GW170817 event. Namely, following the procedure in Refs.~\cite{Altiparmak:2022bke, Jiang:2022tps}, we reject EoSs in the prior ensemble if they do not satisfy any of the following conditions:
\begin{itemize}
\item $M_\text{max} \,\geq\, 2M_\odot$ 
\item $R(M = 2.0M_\odot) \,\geq\, \SI{10.75}{km}$
\item $R(M = 1.1M_\odot) \,\geq\, \SI{10.80}{km}$
\item $\tilde{\Lambda}(\mathcal{M}_\text{chirp} = 1.186 M_\odot) \,\leq\, 720$\end{itemize}
Here, the chirp mass is defined as
\begin{equation}
  \mathcal{M}_\text{chirp} = \frac{(M_1 M_2)^{3/5}}{(M_1 + M_2)^{1/5}}
\end{equation}
for the primary mass $M_1$ and the secondary mass $M_2$, satisfying $M_1 \geq M_2$ in the binary system. The binary tidal deformability $\tilde{\Lambda}$ is expressed as
\begin{equation}
    \tilde{\Lambda} = \frac{16}{13}\frac{(12 M_2 + M_1) M_1^4 \Lambda_1 + (12 M_1 + M_2) M_2^4 \Lambda_2}{(M_1 + M_2)^5}\,,
\end{equation}
which can be regarded as a mass-weighted average of tidal deformability, $\Lambda_{1,2}$ of each star in the binary system. We require the fourth condition for $\tilde{\Lambda}$ for any $M_1$ and $M_2$ that satisfy $M_2 / M_1 > 0.73$. Among the imposed conditions, roughly speaking, the bound on $M_\text{max}$ excludes the bulk of soft EoSs, while the upper bound on $\tilde{\Lambda}$ excludes the bulk of stiff EoSs. We note that the latter was not explicitly incorporated in the EoS construction in the previous section.

We also incorporate the pQCD input as a likelihood, i.e., we reject EoSs with either $P < P_{\min}$ or $P > P_{\max}$ at $\mu = \mu_\text{pQCD}$, where $P_{\min}$ and $P_{\max}$ are $P_\text{pQCD}$ evaluated for $\bar{\Lambda}= \mu$ and $\bar{\Lambda}=4\mu$, respectively. This is effectively equivalent to imposing the QCD likelihood~\cite{Komoltsev:2023zor, Gorda:2022jvk, Gorda:2021znl} at $\mu_\text{pQCD}$. This pQCD input further excludes the stiff EoS in higher-density regions on top of the $\tilde{\Lambda}$ bound. We note that we incorporated the pQCD input in a way that maximizes the impact of the pQCD input onto the inferred EoS results. The impact strongly depends especially on the density up to which the EoS is modeled. See, e.g., Ref.~\cite{Komoltsev:2023zor} for the recent discussion on the constraining power of the pQCD input.

\subsection{Results}

\begin{figure}
    \centering
    \includegraphics[width=.99\columnwidth,clip]{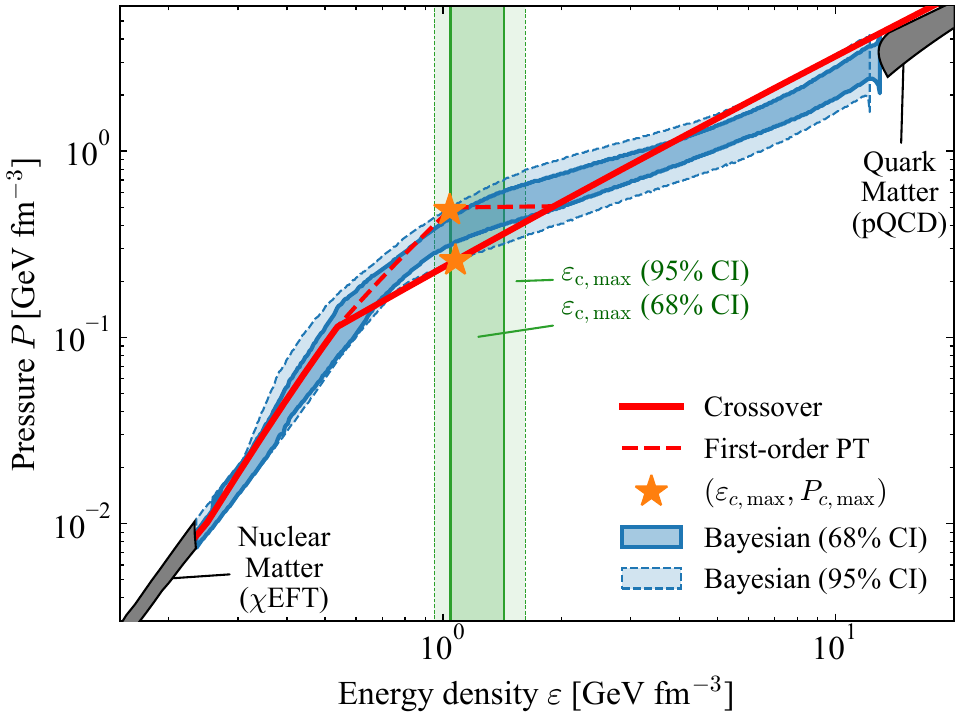}
    \caption{EoSs for the scenarios of ``Crossover'' and ``First-order PT'' overlaid on the Bayesian inference anchoring the pQCD value at high densities. The 68\% and 95\% CIs of the Bayesian results are shown by the solid and dashed curves, respectively. The 68\% and 95\% CIs for the highest energy density at the center of the maximum-mass neutron star are shown by the green bands.}
    \label{fig:bayes_eos}
\end{figure}

The results of our Bayesian inference are presented in Fig.~\ref{fig:bayes_eos}. The posterior distribution is estimated using the histograms of $1.6 \times 10^5$ posterior samples by taking the $500 \times 500$ log-equidistant bins in the $\energy$-$P$ plane within the range $\varepsilon\in [\SI{0.2}{GeV.fm^{-3}}, \SI{14}{GeV.fm^{-3}}]$ and $P\in [\SI{7e-3}{GeV.fm^{-3}}, \SI{4.4}{GeV.fm^{-3}}]$. In Fig.~\ref{fig:bayes_eos}, we plot the 68\% CI (solid curve) and 95\% CI (dashed curve) of the posterior probability distribution.

The posterior distribution shows softening around $\energy \sim \SI{0.7}{GeV.fm^{-3}}$ as was first reported in Ref.~\cite{Annala:2019puf}. This is because the bulk of soft EoSs up to this $\energy$ is excluded by the two-solar-mass constraint. The change in the EoS slope as a whole implies the qualitative change in the underlying degrees of freedom, which we identify as the CO\@. The effective degrees of freedom will be studied further in detail in the next subsection.

The EoSs constructed above for the CO and the first-order PT scenarios are overlaid on top of the Bayesian posteriors in Fig.~\ref{fig:bayes_eos}. It turns out that our CO EoS is the softest among the Bayesian-inferred EoSs near $\varepsilon_{\rm c, \max}$. The 68\% and 95\% CIs for the highest energy density $\energy_{\rm c, \max}$ at the center of the maximum-mass neutron star are shown by the green bands. We see that 68\% of the configurations lie below the onset of the first-order PT in our scenario. This is consistent with our observation that the first-order PT is beyond the stable branch so that this EoS can be effectively regarded as a hadronic EoS.

\subsection{Effective degrees of freedom}

\begin{figure}
    \centering
    \includegraphics[width=.99\columnwidth,clip]{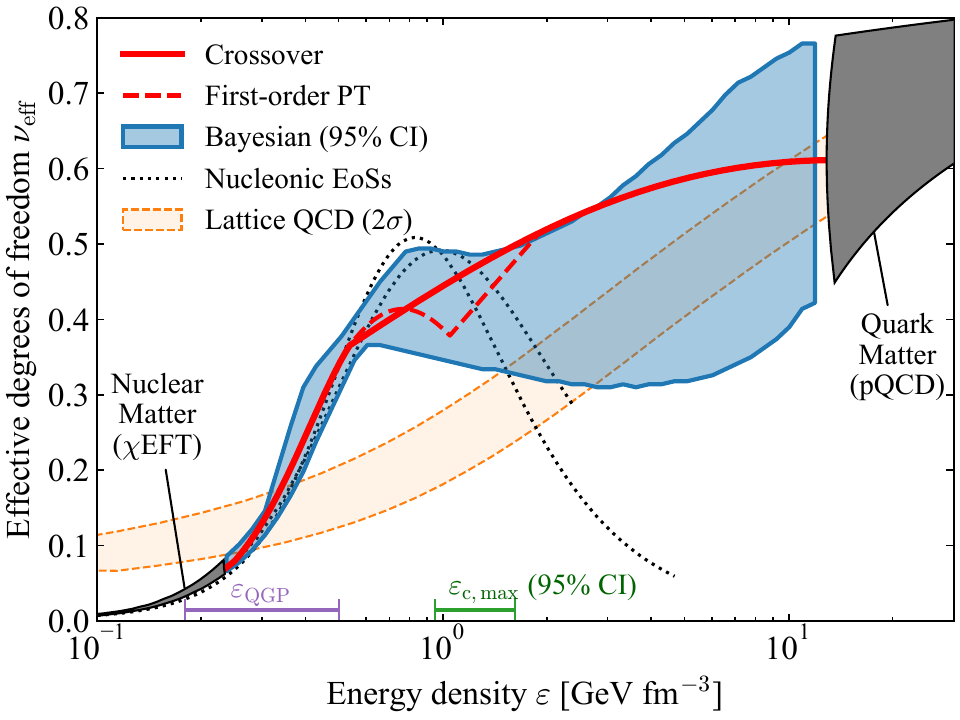}
    \caption{The effective degrees of freedom $\nu_\text{eff}$ for various cold and dense matter EoSs including our CO and PT scenarios (shown by the red solid and dashed lines, respectively), the 95\% CI of the Bayesian results (blue region), and the representative empirical nuclear EoSs~\cite{Akmal:1998cf, Douchin:2001sv} (dotted lines). Also, $\nu_\text{eff}$ from the zero-density and finite-$T$ EoS in the lattice QCD (shown by the orange region) is overlaid.}
    \label{fig:bayes_dof}
\end{figure}

We discuss multiple reasons why quark matter has to be taken into account in matter at high baryon density. To this end, we consider the number of effective degrees of freedom, $\nu_\text{eff}$, which is defined as
\begin{equation}
    \nu_\text{eff} \coloneqq \frac{P}{P_\text{ideal}}\,,
    \qquad
    P_\text{ideal} \coloneqq N_c N_f \frac{\mu^4}{12\pi^2}\,,
    \label{eq:nueff}
\end{equation}
where $P_\text{ideal}$ is the pressure of the ideal quark gas for $N_c=3$ colors and $N_f=3$ flavors, and $\mu$ is the quark chemical potential. We will consider as high density as $\energy \sim \SI{10}{GeV.fm^{-3}}$, so that the strangeness should appear there and the normalization with $N_f=3$ is a natural choice. In our normalization convention, $\nu_\text{eff}\to 1$ in the limit of the ideal quark gas. The purpose of this analysis is to shed light on the underlying physics from the behavior of $\nu_{\rm eff}$ since we expect that this variable may be sensitive to the physical constituents at work at a given energy scale.

Figure~\ref{fig:bayes_dof} presents the 95\% CI of our Bayesian posterior used in Fig.~\ref{fig:bayes_eos} shown in the $\energy$-$\nu_\text{eff}$ plane this time. We take the same $500$ log-equidistant bins as Fig.~\ref{fig:bayes_eos} for $\varepsilon$ and $500$ equidistant bins for $\nu_\text{eff}$ in the range $\nu_\text{eff}\in [0, 2]$. Figure~\ref{fig:bayes_dof} shows the general trend of $\nu_\text{eff}$ monotonically increasing with increasing $\energy$. Up to $\energy \sim \SI{0.7}{GeV.fm^{-3}}$, which corresponds to the point where the sudden change of the slope is observed in Fig.~\ref{fig:bayes_eos}, $\nu_\text{eff}$ increases steeply; beyond this point, $\nu_\text{eff}$ flattens and the increasing rate becomes gradual, or even $\nu_\text{eff}$ decreases a little. Both our CO scenario and the first-order PT scenario capture this feature aptly. The fact that $\nu_\text{eff}$ seems to be saturated around the central density in the maximum-mass neutron star, $\energy_{\rm c, \max}$, may correspond to the appearance of quarks. In the spirit of the conventional (non-interacting) hadron resonance gas, more baryons and resonances contribute to the EoS with increasing density, so $\nu_\text{eff}$ would be expected to increase steadily. By contrast, quarks can only have three flavors at most at this energy scale; therefore, increasing $\nu_\text{eff}$ can naturally be saturated.

The behavior of $\nu_\text{eff}$ can be described from another vantage point based on the conformality of matter~\cite{Fujimoto:2022ohj}. The conformality of matter is quantified by the trace anomaly, $\Theta \coloneqq \energy - 3P$, or by its normalized quantity, $\Delta \coloneqq \Theta / (3\energy)$. Then the increasing rate of $\nu_\text{eff}$ with respect to $\chem$ can be related to the trace anomaly:
\begin{equation}
    \chem \frac{d \nu_{\rm eff}}{d \chem}= \frac{\Theta}{P_\text{ideal}}\,.
\end{equation}
Qualitatively, the slope of $\nu_\text{eff}$ read from Fig.~\ref{fig:bayes_dof} corresponds to the trace anomaly, for $\energy$ is a monotonically increasing function of $\chem$. The monotonically increasing trend of $\nu_\text{eff}$ leads to positive $\Theta$ as conjectured in Ref.~\cite{Fujimoto:2022ohj}, and the nearly flat slope of $\nu_\text{eff}$ corresponds to $\Theta \approx 0$ or $P\approx \energy/3$, i.e., the EoS being approximately conformal. Although there is no rigorous theoretical argument to relate the underlying entity being quark matter with the conformal EoS, it is reasonable to assume that the conformality of matter is associated with quark matter in some form~\cite{Fujimoto:2022ohj, Marczenko:2022jhl, Annala:2023cwx}. We note that such characterization of quark matter does not even need a perturbative regime but can be extended to the strongly-coupled regime. This provides another supportive evidence for our construction of the CO EoS; The adiabatic index is given by $\Gamma = (1 + P/\energy) d \ln P / d \ln\energy$, from which $\Gamma = 4/3$ is obtained for the conformal EoS, $P=\energy/3$, and our choice of $\Gamma$ in our quark matter branch in Table~\ref{tab:pwp} is indeed close to $4/3$.

We also point out a sharp contrast between the EoS incorporating the pQCD effect and the empirical EoSs with the nonrelativistic nucleonic degrees of freedom, for which we use the EoSs from Refs.~\cite{Akmal:1998cf, Douchin:2001sv} as representatives. On the one hand, quark matter described by pQCD, which has a relatively soft EoS close to conformality, bears large degrees of freedom. This is because the asymptotic freedom at high $\energy$ drives the system to approach the ideal quark gas. On the other hand, pure nucleonic matter, which has a stiff EoS in general, has very small degrees of freedom at high $\energy$; see the dotted lines in Fig.~\ref{fig:bayes_dof}. This is understood from the following argument: $P$ is smaller at a given $\chem$ for a stiffer EoS, which is seen also from Eq.~\eqref{eq:Pmu} (see Ref.~\cite{Kojo:2015fua} for further general discussions). From the definition~\eqref{eq:nueff}, thus, $\nu_\text{eff}$ becomes smaller. Because the pQCD EoS is the prediction from the fundamental theory in the Standard Model, $\nu_\text{eff}$ should converge eventually to the pQCD value at energies as high as $\energy \sim \SI{10}{GeV.fm^{-3}}$. This simple observation necessitates changing the degrees of freedom from stiff nuclear matter to soft quark matter.

Finally, let us contrast our high-density CO EoS at $T=0$ with the finite-$T$ QCD EoS of quark-gluon plasma (QGP) at zero density. They reside in different terrains in the phase diagram; nevertheless, they both exhibit CO and the comparison is intriguing. As in Eq.~\eqref{eq:nueff}, we divide $P$ measured on the finite-$T$ lattice~\cite{HotQCD:2014kol} by that of the ideal QCD gas given by
\begin{equation}
    P_\text{ideal} = \left[2(N_c^2 - 1) + \frac{7}{4} (2N_c N_f)\right] \frac{\pi^2 T^4}{90}\,,
\end{equation}
where the first and the second terms in the square bracket are from gluonic and quark degrees of freedom, respectively. In Fig.~\ref{fig:bayes_dof}, finite-$T$ behavior of $\nu_\text{eff}$ is overlaid by the orange region. At first sight, finite-$T$ EoS and finite-density EoS look very different apart from that $\nu_\text{eff}$ is monotonically increasing with growing $\energy$. The dissimilar appearance at low $\energy$ is explained by the difference that the EoS constituents are nonrelativistic for dense and zero-$T$ EoS but relativistic pions are dominant for finite-$T$ EoS\@. Besides, at finite $T$, mesons contribute to the EoS unlike at zero $T$, and heavier degrees of freedom can contribute to the EoS as thermal excitations, while at zero $T$, heavy particles can only enter the EoS once the chemical potential exceeds the mass threshold.

Having said that they appear differently, it is meaningful to compare our CO EoS with the finite-$T$ EoS because QCD has an intrinsic energy scale $\Lambda_\text{QCD}$. Therefore, universal behavior regardless of the difference between $\chem$ and $T$ could be expected for the energy scale corresponding to the transitional region. In Fig.~\ref{fig:bayes_dof}, we show the range of $\energy_\text{QGP}$ corresponding to the CO energy density of high-$T$ QCD matter. Interestingly, this range overlaps with the region where $T=0$ $\nu_\text{eff}$ rapidly increases, and also our CO point is located nearly at the upper edge of this region. If we presume deconfinement of quarks and gluons around $\energy\sim \energy_\text{QGP}$, the purely nucleonic EoS may already be in a CO region and described also as quark matter from a dual nature. See recent discussions in Refs.~\cite{Koch:2024qnz,McLerran:2024rvk} for such a reinterpretation of nuclear matter in the context of quarkyonic duality.

\section{Merger simulation} \label{sec:sim}

Let us turn to dynamical simulations of binary-neutron-star mergers. We start with showing results obtained with our fiducial choice of EoSs, i.e., standard piecewise polytropes and the thermal index $\Gamma_\mathrm{th} = 1.75$ (see below). We will make comparisons with alternative choices of the EoS treatments in Sec.~\ref{sec:sys}.

\subsection{Numerical method and models} \label{sec:sim_method}

\begin{table*}
 \caption{Physical parameters of the binary configurations simulated in this study and characteristic results of simulations. The model name is defined by the primary mass $M_1$ and the secondary mass $M_2$ in the unit of solar mass. The total mass and the mass ratio are denoted by $m_0 \coloneqq M_1 + M_2$ and $q \coloneqq M_2/M_1 (\le 1)$, respectively. The right columns summarize characteristic quantities derived by high-resolution simulations. $M_\mathrm{ej}$ is the mass of the dynamical ejecta and is measured at \SI{10}{\ms} after the first bounce if it occurs or the gravitational collapse otherwise. The definition of these instants are given in Sec.~\ref{sec:sim_lifetime}. The boundness is judged in terms of the Bernoulli criterion. $M_\mathrm{disk}$ is the mass of the disk, i.e., bound material outside the apparent horizon, and is measured only for the case of black-hole formation at \SI{10}{\ms} after the gravitational collapse. $\chi$ is the dimensionless spin parameter of the black hole and is also measured at \SI{10}{\ms} after the gravitational collapse.}
 \label{tab:model}
 \begin{tabular}{ccc|ccp{0.15em}ccp{0.15em}cc}
  \toprule
  Model & $m_0 [M_\odot]$ & $q$ & \multicolumn{2}{c}{$M_\mathrm{ej} [M_\odot]$} && \multicolumn{2}{c}{$M_\mathrm{disk} [M_\odot]$} && \multicolumn{2}{c}{$\chi$} \\
  & & & CO & PT && CO & PT && CO & PT \\
  \midrule
  1.375--1.375 & $2.75$ & $1$     & \num{1.7e-3} & \num{2.3e-3} && 0.054 & --- && $0.71$ & --- \\
  1.4--1.35    & $2.75$ & $0.964$ & \num{3.3e-3} & \num{3.6e-3} && 0.11  & --- && $0.69$ & --- \\
  1.45--1.3    & $2.75$ & $0.897$ & \num{4.1e-3} & \num{3.8e-3} && 0.14  & --- && $0.69$ & --- \\
  1.5--1.25    & $2.75$ & $0.833$ & \num{4.1e-3} & \num{5.1e-3} && 0.15  & --- && $0.68$ & --- \\
  1.55--1.2    & $2.75$ & $0.774$ & \num{4.0e-3} & \num{4.4e-3} && 0.14  & --- && $0.66$ & --- \\
  \midrule
  1.25--1.25 & $2.5$ & $1$ & \num{2.9e-3} & \num{2.9e-3} && --- & --- && --- & --- \\
  1.3--1.3   & $2.6$ & $1$ & \num{2.0e-3} & \num{2.0e-3} && --- & --- && --- & --- \\
  1.35--1.35 & $2.7$ & $1$ & \num{2.5e-3} & \num{2.2e-3} && \num{0.14} & --- && $0.66$ & --- \\
  1.4--1.4   & $2.8$ & $1$ & \num{3.4e-3} & \num{3.2e-3} && 0.042 & --- && $0.74$ & --- \\
  1.45--1.45 & $2.9$ & $1$ & \num{7.6e-4} & \num{4.7e-3} && \num{8.6e-3} & --- && $0.78$ & --- \\
  1.5--1.5   & $3.0$ & $1$ & \num{3.0e-4} & \num{1.5e-2} && \num{1.3e-4} & --- && $0.82$ & --- \\
  1.55--1.55 & $3.1$ & $1$ & \num{2.e-5} & \num{5.0e-4} && \num{2.e-5} & \num{2.1e-4} && $0.81$ & $0.80$ \\
  1.6--1.6   & $3.2$ & $1$ & $<\num{e-5}$ & $<\num{e-5}$ && $<\num{e-5}$ & $<\num{e-5}$ && $0.80$ & $0.80$ \\
  1.65--1.65 & $3.3$ & $1$ & $<\num{e-5}$ & $<\num{e-5}$ && $<\num{e-5}$ & $<\num{e-5}$ && $0.79$ & $0.79$ \\
  1.7--1.7   & $3.4$ & $1$ & $<\num{e-5}$ & $<\num{e-5}$ && $<\num{e-5}$ & $<\num{e-5}$ && $0.78$ & $0.78$ \\
  \bottomrule
 \end{tabular}
\end{table*}

Dynamical simulations of the coalescence are performed with a well-tested numerical-relativity code, {\small SACRA} \cite{Yamamoto:2008js}. The Einstein evolution equation is solved by the puncture-BSSN formulation \cite{Shibata:1995we,Baumgarte:1998te,Campanelli:2005dd,Baker:2005vv,Marronetti:2007wz} with locally implementing Z4c prescriptions for constraint propagation~\cite{Bernuzzi:2009ex,Hilditch:2012fp}. The details are described in Ref.~\cite{Kyutoku:2014yba} and references therein.

Both zero- and finite-$T$ parts of the EoS need to be specified in merger simulations because the collision of two neutron stars gives rise to the temperature $T \sim \order{10}\si{\mega\eV}$. In this study, the zero-$T$ part is given according to the assumptions on the hadron-quark transition, i.e., the CO or the first-order PT, as described in Sec.~\ref{sec:eos}. The finite-$T$ effect is incorporated by adopting the so-called $\Gamma_\mathrm{th}$ prescription, in which the thermal pressure, $P_\mathrm{th}$, is assumed to take an ideal-gas-like form as a function of the finite-$T$ contribution to the internal energy $\varepsilon_\mathrm{th}$. Specifically, we assume the following form of the finite-$T$ EoS,
\begin{equation}
    P_\mathrm{th} (\rho , \varepsilon) = (\Gamma_\mathrm{th} - 1) \varepsilon_\mathrm{th}\,,
\end{equation}
where
\begin{equation}
    \varepsilon_\mathrm{th} (\rho , \varepsilon) \coloneqq \varepsilon - \varepsilon_\mathrm{cold} (\rho)
\end{equation}
is obtained in numerical simulations with the aid of the assumed EoS at $T=0$. We adopt $1.75$ as the fiducial value of the constant parameter in this prescription, $\Gamma_\mathrm{th}$ \cite{Bauswein:2010dn,Figura:2020fkj}. The total pressure is given by $P = P_\mathrm{cold} + P_\mathrm{th}$. We will discuss systematic uncertainties associated with these implementations in Sec.~\ref{sec:sys}.

Initial data of simulations are given by quasiequilibrium states of irrotational binary neutron stars~\cite{Gourgoulhon:2000nn,Kyutoku:2014yba}. Physical parameters of the models employed in this work are shown in Table~\ref{tab:model}. All the models are prepared at the orbital angular velocity of $\Omega_0 \approx 0.025 /(Gm_0)$, where $m_0$ is the total mass of the system, with the orbital eccentricity reduced to $\sim 0.1\%$~\cite{Kyutoku:2014yba}. Because the maximum rest-mass density does not reach $\rho_3$ of our EoS in any initial configurations, the initial data are the same for the CO and PT scenarios. Actual computations of quasiequilibrium states are performed using the spectral method library, {\small LORENE} \cite{2016ascl.soft08018G}.

We employ at least two grid resolutions for all the models to check the truncation errors. Specifically, all the models are simulated with the high resolution of $\approx \SI{190}{\meter}$ and the low resolution of $\approx \SI{280}{\meter}$ in the region around the neutron stars. For selected models, we also employ the ultrahigh resolution of $\approx \SI{140}{\meter}$ and the middle resolution of $\approx \SI{230}{\meter}$.\footnote{To be exact, the grid spacing changes by the ratio $3:4:5:6$ for the ultrahigh, high, middle, and low resolutions.} By employing an adaptive mesh refinement algorithm (see Ref.~\cite{Yamamoto:2008js} for details), the outer boundaries of computational domains are located at $\approx \SI{3000}{\km}$ from an approximate center of mass of the system with ten refinement levels. Specifically, two sets of four-leveled finer resolution grids comove with two neutron stars, and these grids are covered by six-leveled coarser resolution grids. The material that left the computational domain is monitored and taken into account appropriately when estimating the ejected mass. Finally, in all the simulations, we impose the reflection symmetry with respect to the orbital plane.

Our numerical simulations are continued for $\approx \SI{30}{\ms}$ after two neutron stars merge. This timescale is chosen partly because we solve neither the magnetic field nor the neutrino transport. As these physical effects will play an important role at late times~\cite{Hotokezaka:2013iia}, i.e., $\gtrsim \SI{100}{\ms}$ for the magnetic field and $\gtrsim \SI{1}{\second}$ for the neutrino transport, simulations longer than $\approx \SI{30}{\ms}$ will not give realistic results. Fortunately, gravitational-wave emission becomes very weak after $\approx \SI{30}{\ms}$. Thus, our present setup for simulations is sufficient to judge the identifiability of the physical transition scenario via gravitational-wave observations. Quantitative measures of the identifiability are discussed along the realistic process of data analysis in a separate study~\cite{Harada:2023eyg}.

\subsection{Merger dynamics} \label{sec:sim_dynamics}

\begin{figure*}
    \begin{tabular}{c}
    \includegraphics[width=.95\linewidth,clip]{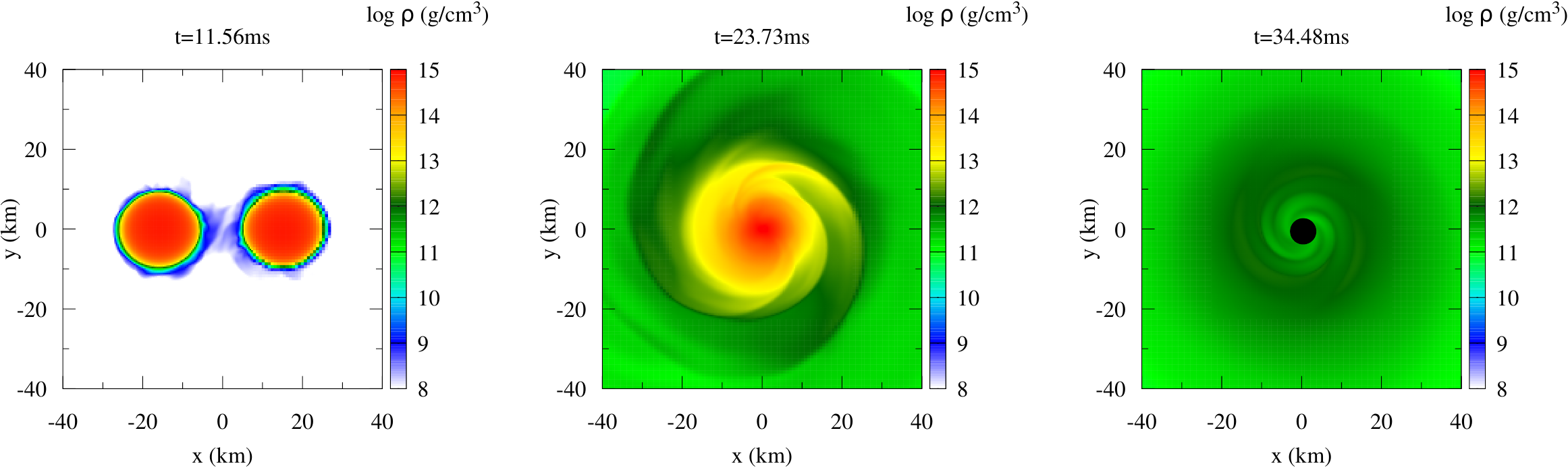} \\
    \includegraphics[width=.95\linewidth,clip]{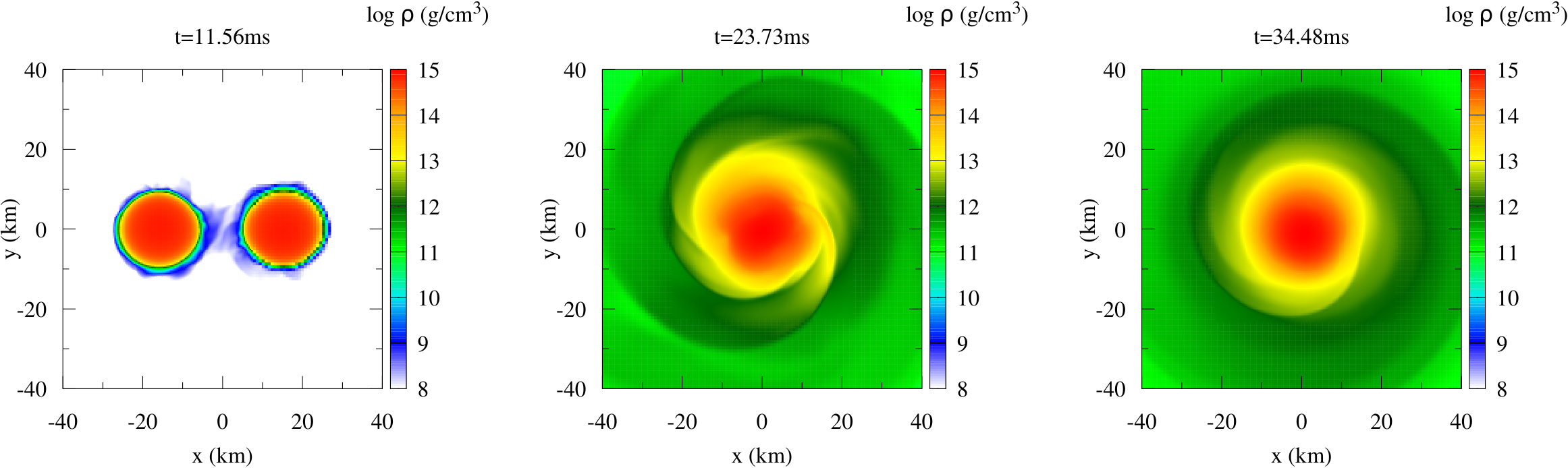}
    \end{tabular}
    \caption{Snapshots of the rest-mass density on the orbital plane at $t - t_\mathrm{fb} \approx \SI{-3}{\ms}$ (left), $\approx \SI{9}{\ms}$ (middle), and $\approx \SI{20}{\ms}$ (right) of the model 1.4--1.35 for the CO (top) and PT (bottom) scenarios. Here, $t_\mathrm{fb}$ denotes the time of the first bounce defined in Sec.~\ref{sec:sim_lifetime}. The filled black circle on the top right panel denotes the interior of the apparent horizon.} \label{fig:snapshot}
\end{figure*}

As a representative case, Fig.~\ref{fig:snapshot} shows the snapshots of the rest-mass density on the orbital plane for the model 1.4--1.35 in the CO (top panels) and PT (bottom panels) scenarios. The dynamics are identical during the inspiral phase because the EoSs are the same for the CO and PT scenarios up to the maximum density inside the premerger neutron stars. This situation holds for all the models listed in Table~\ref{tab:model}, where the most massive (premerger) neutron star is $1.7M_\odot$. This indicates that the inspiral gravitational waveforms, while beneficial for constraining the EoS in the intermediate-density range, will not be useful for distinguishing our transition scenarios unless a neutron star with $\approx 2M_\odot$ is involved in the merger (see Ref.~\cite{Kyutoku:2020xka} for an extensive discussion of GW190425-like systems). Gravitational-wave signals will be discussed in Sec.~\ref{sec:sim_gw}.

Because the total mass of the system for this model, $2.75M_\odot$ motivated by GW170817, is not extremely high, the binary coalescence results in the formation of a massive remnant neutron star~\cite{Hotokezaka:2011dh}. In addition, the high adiabatic index of $\Gamma_3 = 3.5$ tends to keep the merger remnant nonaxisymmetric~\cite{Shibata:2005ss}, further preventing immediate collapse. The high degree of nonaxisymmetry helps the angular momentum be redistributed from the central core to the outer envelope via the hydrodynamic torque and also be emitted by gravitational radiation. As a result of losing centrifugal support, the maximum rest-mass density of the system gradually increases with time (see discussions around Fig.~\ref{fig:rhomax}).

The transition scenarios have a critical impact on the fate of the postmerger remnant. In the CO scenario, the EoS becomes soft as the pQCD branch is approached, and the gravitational collapse sets in to form a black hole at $\approx \SI{10}{\ms}$ after the collision of neutron stars. By contrast, in the first-order PT scenario, for which the total mass of the system $m_0$ is smaller than the typical maximum mass of supramassive neutron stars, $\approx 1.2M_\mathrm{max}$ (see, e.g., Refs.~\cite{Cook:1993qr,Morrison:2004fp,Breu:2016ufb}), the remnant does not collapse into a black hole in our simulation. As we will see in Sec.~\ref{sec:sim_gw}, these two scenarios can be distinguished by whether the postmerger gravitational-wave signal terminates abruptly or not.

\subsection{Gravitational waves} \label{sec:sim_gw}

\begin{figure*}
    \begin{tabular}{cc}
    \includegraphics[width=.48\linewidth,clip]{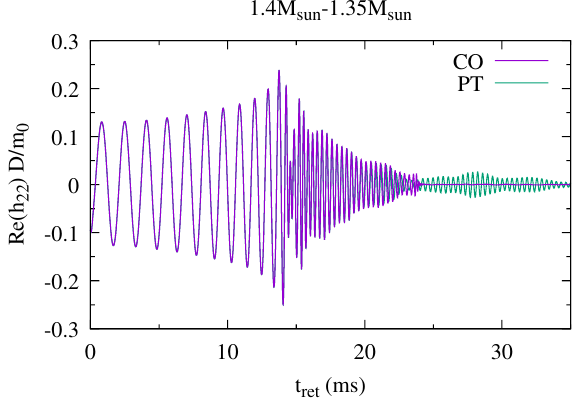} &
    \includegraphics[width=.48\linewidth,clip]{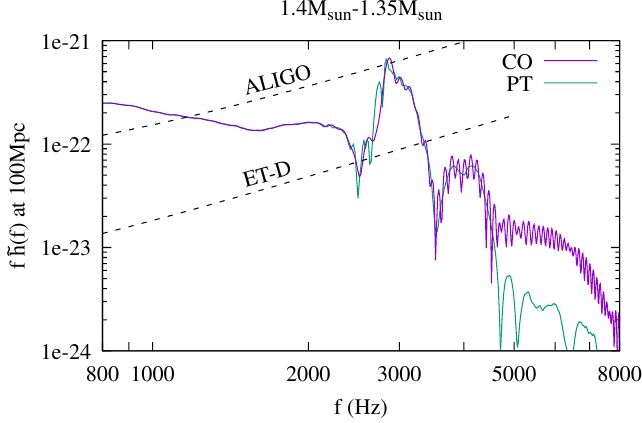}
    \end{tabular}
    \caption{Gravitational waveform (left) and spectrum (right) for the model 1.4--1.35. The purple and green curves show the results for the CO and PT scenarios, respectively. The spectrum is presented for a hypothetical distance of \SI{100}{Mpc} and compared with the design sensitivity of Advanced LIGO (ALIGO) and Einstein Telescope (ET-D) \cite{LIGOScientific:2016wof}. Wiggles on the spectrum for the CO scenario is caused by the minor amplitude peak at the moment of black-hole formation.} \label{fig:gw}
\end{figure*}

As we have stated above, postmerger gravitational waves serve as the most direct tool for distinguishing the two transition scenarios.

\subsubsection{Case study: 1.4--1.35}

The left panel of Fig.~\ref{fig:gw} compares the gravitational waveforms and spectra for the model 1.4--1.35 depicted in Fig.~\ref{fig:snapshot} (see also Ref.~\cite{Fujimoto:2022xhv} for spectrograms of the models 1.375--1.375 and 1.55--1.2). Because the dynamics are identical during the inspiral phase, the gravitational waveforms are also the same. The gravitational-wave amplitude peaks approximately when the two neutron stars collide \cite{Kiuchi:2017pte}, and soon after this moment, gravitational waves for the two transition scenarios begin to deviate reflecting the different high-density EoSs. The most remarkable difference is the sudden shutdown of the signal for the CO scenario at $\approx \SI{10}{\ms}$ after the peak. This signifies the gravitational collapse of the remnant into a black hole. Indeed, detailed investigations reveal that the end of the shutdown is characterized by the quasinormal mode of the remnant black hole (see, e.g., Refs.~\cite{Baiotti:2008ra,Kiuchi:2009jt} for relevant earlier studies).

The gravitational-wave spectrum shown in the right panel of Fig.~\ref{fig:gw} also clearly shows the distinct behavior between the two scenarios in a different way than the time-domain waveform. If the merger results in forming a massive neutron star at least transiently, a prominent peak appears at $\approx \SI{3}{\kilo\hertz}$, which is mainly determined by the dynamical timescale of the neutron star. A close look reveals that the peak frequency for the CO scenario is slightly higher (quantitatively by 1\%--2\%) than that for the PT scenario. The high-frequency components for the former are caused by the slightly compact remnant associated with the softening toward the pQCD branch. Although the lifetime is different, the amplitude at the spectral peak is approximately the same for the two scenarios. This is because the postmerger gravitational emission is not very strong for $\gtrsim \SI{10}{\ms}$ after the collision. These findings are consistent with the previous simulations for non-strong PT (e.g., Ref.~\cite{Bauswein:2018bma}).

To distinguish these two scenarios, it is indispensable to observe high frequency $\gtrsim \SI{2}{\kilo\hertz}$ corresponding to postmerger emission with sufficient sensitivity. This high-frequency range is hardly accessible with current gravitational-wave detectors \cite{LIGOScientific:2017fdd}, and third-generation and/or specially-designed detectors will play a pivotal role. In particular, if the remnant collapses early into a black hole as in our CO scenario, the spectrum extends with a moderate amplitude up to a high cutoff frequency of $\approx \SI{7}{\kilo\hertz}$ associated with quasinormal modes of the remnant black hole \cite{Dhani:2023ijt}. Ultimately, identification of the quasinormal-mode excitation will strongly indicate black-hole formation, and in our scenario, it is interpreted as the onset of quark matter in the CO scenario. Unfortunately, the quasinormal-mode cutoff itself is still far below the sensitivity of third-generation detectors for \SI{100}{Mpc} as shown in the right panel of Fig.~\ref{fig:gw}. Nevertheless, these detectors will in principle enable us to distinguish two scenarios if the matched-filtering analysis can be performed with the precise theoretical templates of gravitational waves including the phase \cite{Harada:2023eyg} (see also Refs.~\cite{Easter:2021wlb,Breschi:2022ens,Dhani:2023ijt,Prakash:2023afe} for related studies).

\subsubsection{Parameter dependence}

\begin{figure*}
    \begin{tabular}{cc}
         \includegraphics[width=.48\linewidth,clip]{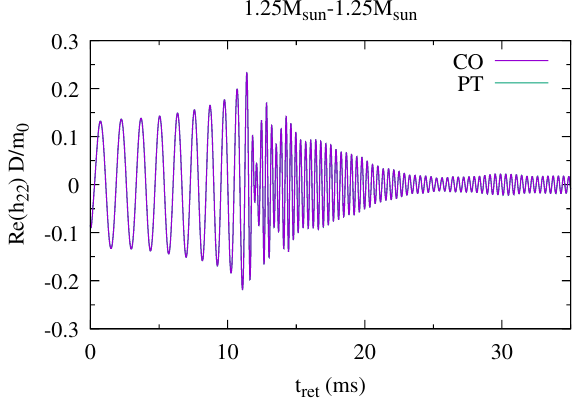} &
         \includegraphics[width=.48\linewidth,clip]{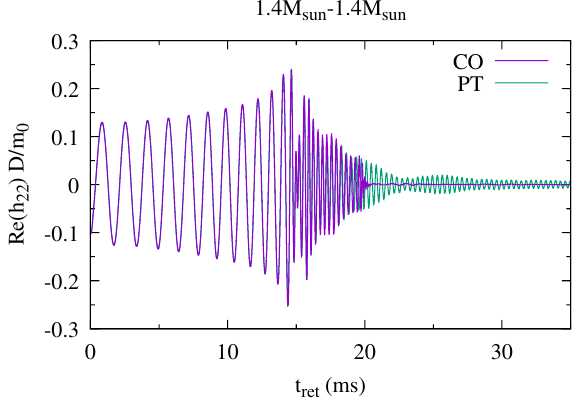} \\
         \includegraphics[width=.48\linewidth,clip]{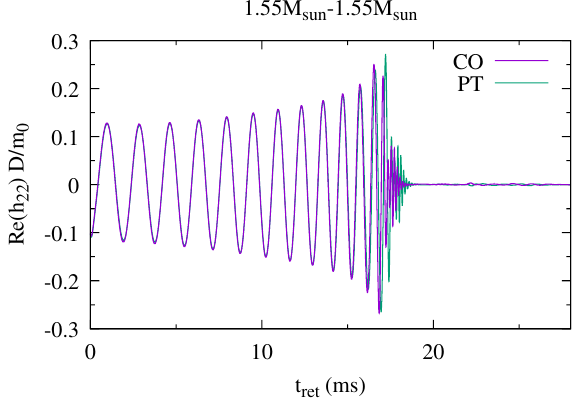} &
         \includegraphics[width=.48\linewidth,clip]{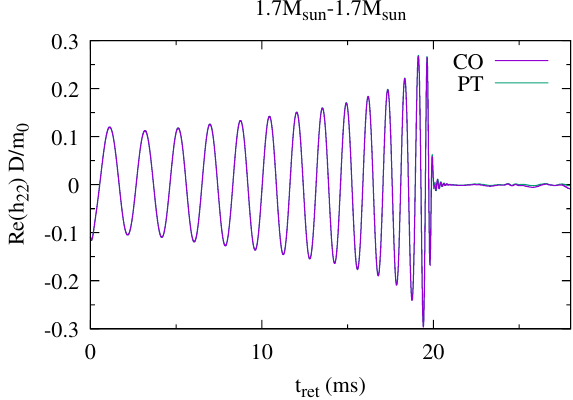}
    \end{tabular}
    \caption{Same as the left panel of Fig.~\ref{fig:gw} but for the models 1.25--1.25 (top left), 1.4--1.4 (top right), 1.55--1.55 (left bottom), and 1.7--1.7 (right bottom).} \label{fig:gw_mtot}
\end{figure*}

The total mass of the system, $m_0$, is the key factor for determining the fate of the binary coalescence. Figure~\ref{fig:gw_mtot} compares the gravitational waveforms from various equal-mass systems with different values of $m_0$. The lightest model 1.25--1.25 results in the formation of a long-lived remnant irrespective of the transition scenario, making the distinction impossibly difficult. As the total mass increases, e.g., 1.4--1.4, the remnant for only the CO scenario collapses early. This situation is qualitatively the same as the case in the model 1.4--1.35 shown in Fig.~\ref{fig:gw}. The difference between the two scenarios is the most prominent for these ranges of $m_0$. It is quite encouraging that this informative mass range overlaps with astrophysical populations of binary neutron stars \cite{Farrow:2019xnc} including GW170817. If $m_0$ is as large as 1.55--1.55, the binary merger results in the prompt collapse for both the CO and PT scenarios. Accordingly, the difference is minor even in the postmerger waveforms. For a very massive system like 1.7--1.7, gravitational waves shut down immediately after merger for both scenarios, again rendering the distinction of scenarios difficult. These comparisons suggest that there should exist a useful range of the total mass, $m_0$, in the middle of the realistic distribution. We will quantify this information in Sec.~\ref{sec:sim_lifetime}.

\begin{figure*}
    \begin{tabular}{cc}
         \includegraphics[width=.48\linewidth,clip]{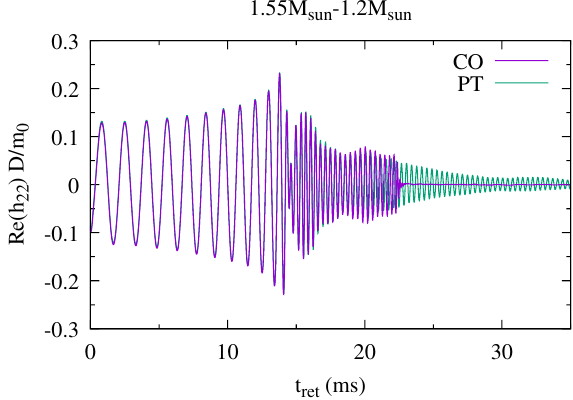} &
         \includegraphics[width=.48\linewidth,clip]{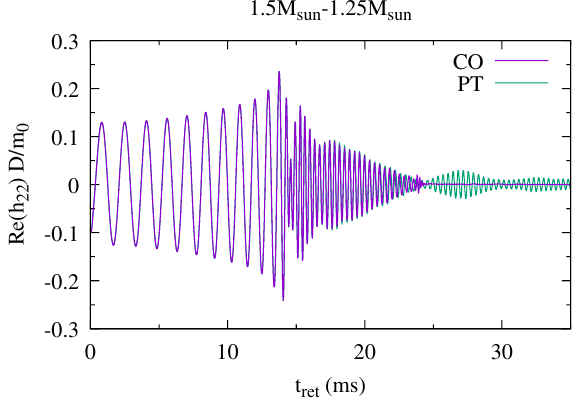} \\
         \includegraphics[width=.48\linewidth,clip]{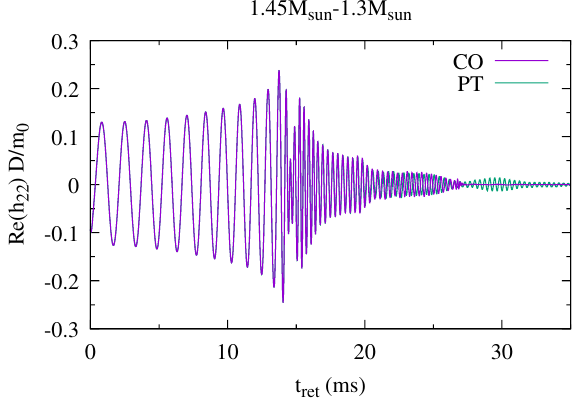} &
         \includegraphics[width=.48\linewidth,clip]{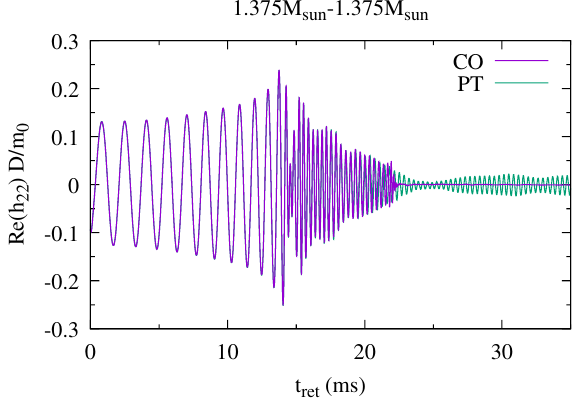}
    \end{tabular}
    \caption{Same as the left panel of Fig.~\ref{fig:gw} but for the models $1.2$--$1.55$ (top left), $1.25$--$1.5$ (top right), $1.3$--$1.45$ (left bottom), and $1.375$--$1.375$ (right bottom).} \label{fig:gw_q}
\end{figure*}

We find that the difference between the gravitational-wave signals for the two transition scenarios depends only weakly on the mass ratio, $q$. Figure \ref{fig:gw_q} compares the gravitational waveforms with different values of $q$ for a fixed choice of $m_0 = 2.75M_\odot$. It is evident that, irrespective of $q$ for the CO scenario, the sudden shutdown associated with the gravitational collapse occurs at $\approx \SI{10}{\milli\second}$ after the first bounce. Similarly, the PT scenario always results in the formation of long-lived remnants due to the large maximum mass. Such weak dependence on $q$ is a desirable feature, for gravitational-wave observations can infer the mass ratio only with a moderate precision~\cite{LIGOScientific:2017vwq,LIGOScientific:2020aai}. On another front, the mass ratio will have a sizable impact on the electromagnetic emission as addressed in Sec.~\ref{sec:sim_mass}.

\subsection{Lifetime of the remnant} \label{sec:sim_lifetime}

The distinguishability of the two transition scenarios may be characterized semiquantitatively by the lifetime of the remnant, hereafter denoted by $t_\mathrm{life}$. Precisely speaking, we define $t_\mathrm{life}$ by the duration from the first bounce, i.e., the first local maximum in time of the maximum rest-mass density in space, to the gravitational collapse, i.e., the first local minimum in time of the minimum (gauge-dependent) lapse function around $0$ in space.\footnote{Another reasonable definition of the gravitational collapse may be the apparent-horizon formation. This changes $t_\mathrm{life}$ only by $<\SI{1}{\ms}$.} We only put a lower limit on the values of $t_\mathrm{life}$ if it exceeds $\SI{30}{\ms}$. We recall that unmodeled physics such as magnetohydrodynamics plays a significant role after this time in reality \cite{Hotokezaka:2013iia}. This limitation is unlikely to affect our main conclusion about the distinguishability with gravitational waves, because the emission is weak after this epoch. Detailed discussions of the distinguishability is presented in a separate work~\cite{Harada:2023eyg}.

\begin{figure*}
    \begin{tabular}{cc}
    \includegraphics[width=.48\linewidth,clip]{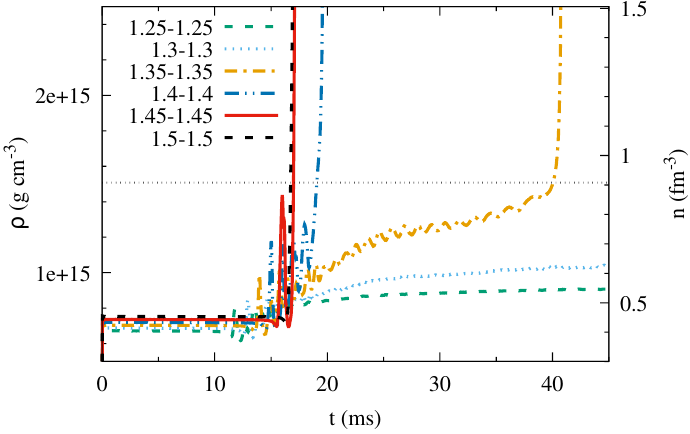} &
    \includegraphics[width=.48\linewidth,clip]{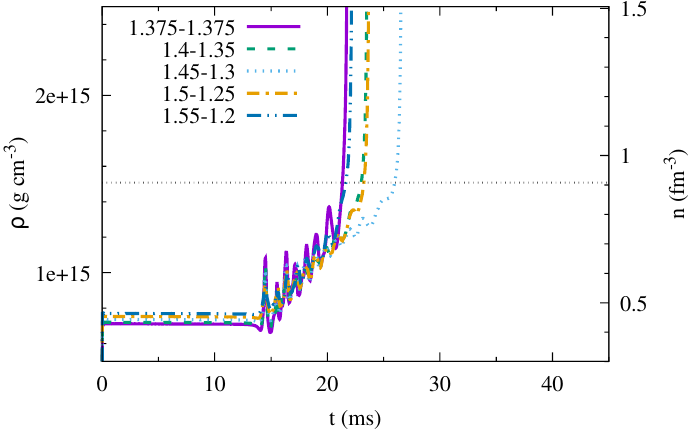}
    \end{tabular}
    \caption{Evolution of the maximum rest-mass density of the system for various models in the CO scenario. The left panel shows the results for the equal-mass models with different values of $m_0$. The right panel shows the results for the models with $m_0 = 2.75M_\odot$ with different values of $q$. The dotted line indicates the rest-mass density at the center of a maximum-mass spherical neutron star, $\approx \SI{1.5e15}{\gram\per\cubic\cm}$.} \label{fig:rhomax}
\end{figure*}

We find that the gravitational collapse for the CO scenario sets in approximately when the maximum rest-mass density of the neutron star reaches that of the maximum-mass spherical neutron star, that is $\approx \SI{1.5e15}{\gram\per\cubic\cm}$. Figure~\ref{fig:rhomax} shows the time evolution of the maximum rest-mass density of the system for various models in the CO scenario. Once the maximum rest-mass density exceeds this threshold, it begins to increase rapidly on a dynamical timescale of $\lesssim \si{\ms}$ toward the black-hole formation. This feature suggests that compact binary coalescences will enable us to explore approximately the same density range as the stable spherical neutron stars do, consistently with previous findings \cite{Ujevic:2023vmo}. It may be worthwhile to comment that the genuine maximum mass of a spherical neutron star might be reached only in transient events such as binary-neutron-star mergers.

For the PT scenario, the lifetime drastically changes from $t_\mathrm{life} \ge \SI{30}{\ms}$ at $m_0 = 3.0M_\odot$ to $\approx \SI{0}{\ms}$ at $3.1M_\odot$. From the evolution of the maximum rest-mass density, we have been unable to identify the critical density even approximately. This is simply because our choices of $m_0$ are not sufficiently fine to resolve the transition region. This could be tackled in a straightforward manner by performing further simulations in this region, and we leave this for future work.

\begin{figure*}
    \begin{tabular}{cc}
    \includegraphics[width=.48\linewidth,clip]{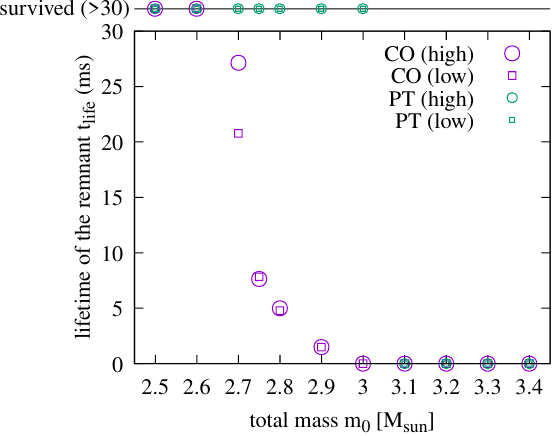} &
    \includegraphics[width=.48\linewidth,clip]{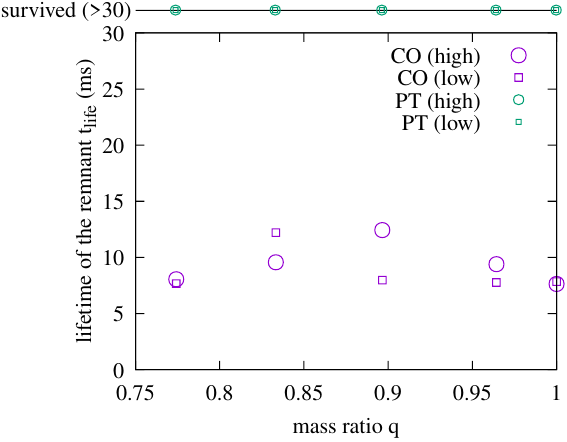}
    \end{tabular}
    \caption{Lifetime of the remnant, $t_\mathrm{life}$, as a function of the total mass of the system, $m_0$, for equal-mass ($q=1$) models (left) and $t_\mathrm{life}$ as a function of the mass ratio, $q$, for $m_0 = 2.75M_\odot$ models (right). The large-purple and small-green symbols represent the results for the CO and PT scenarios, respectively. Results for high ($\approx \SI{190}{\meter}$) and low ($\approx \SI{280}{\meter}$) resolutions are shown by circles and squares, respectively, to check numerical uncertainties.} \label{fig:life}
\end{figure*}

The left panel of Fig.~\ref{fig:life} shows the lifetime of the merger remnant, $t_\mathrm{life}$, as a function of the total mass, $m_0$, for equal-mass ($q=1$) models. This figure confirms the indication of Fig.~\ref{fig:gw_mtot} that $t_\mathrm{life}$ depends primarily on $m_0$ in a quantitative manner. For the CO scenario, the merger outcome is formation of a long-lived remnant with $t_\mathrm{life} > \SI{30}{\ms}$ for $m_0 < 2.7M_\odot$. On the opposite side, the prompt collapse is observed for $m_0 > 2.9M_\odot$. The transitional change between these two cases occurs gradually in the range $m_0 = 2.7M_\odot$ to $2.9M_\odot$. This mass range suggests that future gravitational-wave detection of multiple binary-neutron-star mergers with various total masses may enable us to characterize the nature of possible QCD CO in a detailed manner. For example, the remnant of GW170817 with $m_0\approx 2.75M_\odot$ survives $\approx 5$--$\SI{10}{\ms}$ in our CO scenario, while GW190425 with $m_0\approx 3.4M_\odot$ promptly forms a black hole. Galactic binary neutron stars have the total masses of $m_0\approx 2.5$--$2.9M_\odot$~\cite{Farrow:2019xnc}, and thus their mergers will be useful for observationally clarifying the dependence of the lifetime on the total mass. By compiling future postmerger gravitational-wave signals from these systems, we may be able to reveal the detailed EoS behavior in the transition region.

The lifetime in our PT scenario also exhibits a change between $m_0 = 3.0M_\odot$ and $3.1M_\odot$ as described above. Taking the fact that GW170817 and GW190425 enclose this mass range into account, gravitational-wave observations might constrain the EoS also in this scenario. However, we caution that our EoS for this PT scenario is manually constructed to avoid the superluminal sound speed. Thus, the precise value of the threshold mass cannot be predicted from theory. Accordingly, we refrain from going into detailed discussions of its possible implications. Yet, we conclude a robust lesson that, if the lifetime turns out to be $t_\mathrm{life} > \SI{30}{\ms}$ even for $m_0 \approx 3.0M_\odot$, our CO scenario would be altered.

This figure also compares the results derived with high and low resolutions. The variation of $t_\mathrm{life}$ is only $\lesssim 20\%$ for the range considered in this study. When available, ultrahigh and middle resolutions also derive the results within the similar uncertainty range. Although the convergence is not very clean due to the metastable nature of hypermassive neutron stars, particularly for configurations near the threshold of prompt collapse (see Refs.~\cite{Hotokezaka:2011dh,Bauswein:2013jpa,Koppel:2019pys} for relevent studies), this finding confirms that the truncation errors have only a limited impact on our current discussions about distinguishability.

The right panel of Fig.~\ref{fig:life} shows $t_\mathrm{life}$ as a fucntion of the mass ratio, $q$, for $m_0 = 2.75M_\odot$ models. Consistently with Fig.~\ref{fig:gw_q}, the dependence is weak and nonmonotonic. These features have been found for the majority, if not all, of EoSs in previous work~\cite{Hotokezaka:2013iia}. Although the dependence on the grid resolution is also weak, we find it comparable to the variation caused by the mass ratio. Significantly high resolutions should be necessary for quantifying the physical dependence on the mass ratio.

\subsection{Remnant disk and ejected material} \label{sec:sim_mass}

Electromagnetic counterparts to compact binary mergers also provide us with valuable information about the merger dynamics and the underlying EoS properties. Although the detailed prediction of electromagnetic signals is out of our current scope, the masses of the remnant disk and the ejected material serve as a basis for future considerations. In particular, because the kilonova AT\;2017gfo, that followed GW170817, is likely to require massive ejecta of $\approx 0.05M_\odot$~\cite{Kasen:2018drm,Hotokezaka:2019uwo}, the realistic EoS must leave this amount of material in any form outside the black hole at the very least. In reality, substantially larger masses will be required, because the ejection efficiency from the remnant disk is likely to be $\order{10\%}$ (see, e.g., Refs.~\cite{Fernandez:2013tya,Fujibayashi:2020qda,Kiuchi:2022nin}).

In this study, we estimate the amount of unbound material based on the Bernoulli criterion. That is, the material outside the apparent horizon is regarded as unbound if $-hu_t >1$, and vice versa, where $u_t$ is the covariant time component of the four velocity. It should be cautioned that this criterion could overestimate the amount of mass ejection, because the internal energy is not necessarily converted to the kinetic energy before the fallback. On another front, the geodesic criterion based on $-u_t$ rather than $-hu_t$ inevitably underestimates the unbound mass (see Refs.~\cite{Foucart:2021ikp,Fujibayashi:2022ftg} for further discussions). Precise evaluation requires long-term simulations incorporating the \textit{r}-process heating \cite{Desai:2018rbc,Kawaguchi:2020vbf,Ishizaki:2021qne}. Fortunately, this uncertainty does not affect our discussion about AT\;2017gfo, because it turns out later that at most $\approx \num{5e-3}M_\odot$ will become unbound in our simulations. Because no physical mechanisms for launching the disk outflow is incorporated in this study, this unbound material should be considered to represent only the dynamical ejecta. While its modeling is beyond the scope of this study, the disk outflow is essential to explain AT\;2017gfo \cite{Kasen:2017sxr,Tanaka:2017qxj}.

\subsubsection{Numerical result}

Table~\ref{tab:model} lists the bound mass $M_\mathrm{disk}$ and the unbound mass $M_\mathrm{ej}$. As implied by the subscript, $M_\mathrm{disk}$ and $M_\mathrm{ej}$ are regarded as the disk and the ejecta mass, respectively, in this study. The former is presented only for the case of black-hole formation and measured at \SI{10}{\ms} after the gravitational collapse. The latter is measured at \SI{10}{\ms} after the first bounce if it occurs. Otherwise, it is also measured at \SI{10}{\ms} after the gravitational collapse.

\begin{figure*}
    \begin{tabular}{cc}
    \includegraphics[width=.48\linewidth,clip]{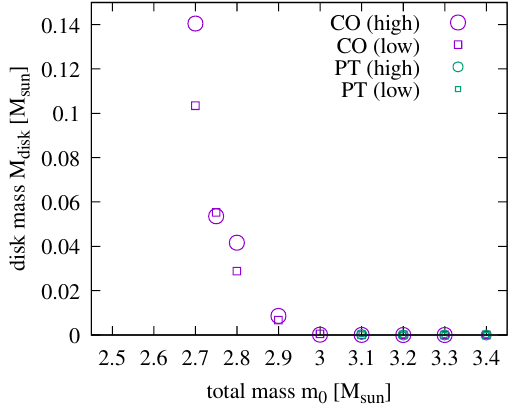} &
    \includegraphics[width=.48\linewidth,clip]{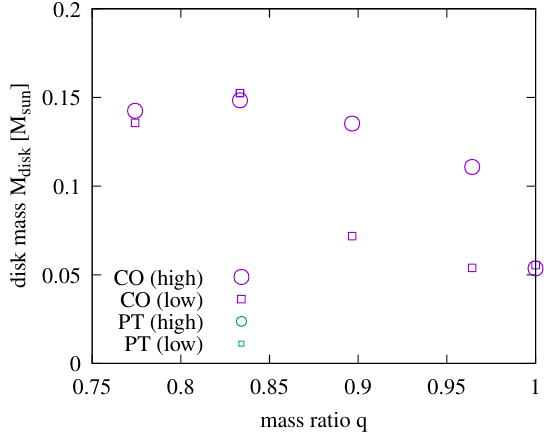}
    \end{tabular}
    \caption{Mass of the remnant disk, $M_\mathrm{disk}$, as a function of the total mass of the system, $m_0$, for equal-mass models (left) and as a function of the mass ratio, $q$, for $m_0 = 2.75M_\odot$ models (right). The large purple and small green symbols represent the results for the CO and PT scenarios, respectively. Results for high ($\approx \SI{190}{\meter}$) and low ($\approx \SI{280}{\meter}$) resolutions are shown with circles and squares, respectively, to check numerical uncertainties.} \label{fig:mdisk}
\end{figure*}

Figure \ref{fig:mdisk} visualizes the dependence of $M_\mathrm{disk}$ on the total mass, $m_0$, and the mass ratio, $q$. We do not plot the points for the (trivial) cases with $m_0 < 2.7M_\odot$ for CO and $< 3.1M_\odot$ for PT, in which remnant massive neutron stars survive for \SI{30}{\ms} after the first bounce. To give an idea of truncation errors, we also present the results of low-resolution simulations as in Fig.~\ref{fig:life}. Here, in Fig.~\ref{fig:mdisk}, both $m_0$ and $q$ are found to play an important role.

The dependence of $M_\mathrm{disk}$ on $m_0$ shown in the left panel is essentially the same as that of $t_\mathrm{life}$. That is, the disk mass is larger for a lower-mass system with a longer lifetime of the remnant massive neutron star. This is because the angular momentum transferred to the outer part of the remnant increases as $t_\mathrm{life}$ increases, resulting in a massive disk after the black-hole formation. Indeed, a comparison with Fig.~\ref{fig:life} reveals that the effect of the grid resolution on $M_\mathrm{disk}$ largely follows the variation of $t_\mathrm{life}$ in accordance with the consideration above. If the black hole is formed promptly as in the cases of $m_0 > 3.0M_\odot$ for both scenarios, essentially no material is left outside, rendering electromagnetic emission unlikely.

The right panel of Fig.~\ref{fig:mdisk} shows that, differently from the case of $t_\mathrm{life}$ in Fig.~\ref{fig:life}, $M_\mathrm{disk}$ increases significantly with the decrease of $q$ \cite{Hotokezaka:2012ze,Kiuchi:2019lls}. Here, the large amount of $M_\mathrm{disk}$ in asymmetric systems with small values of $q$ is attributed to the high degree of tidal deformation of the secondary prior to merger. The tidal torque exerted on the deformed structure enhances the angular momentum transport and thus the disk formation. Remarkably, even mild mass asymmetry, say $q = 0.964$, more than doubles the value of $M_\mathrm{disk}$ compared to the equal-mass case. For a given value of $q$, the effect of the grid resolution is again reasonably understood in view of the variation of $t_\mathrm{life}$. Quantitatively, the grid resolution affects the value of $M_\mathrm{disk}$ by a factor of order unity for only mildly asymmetric systems with $0.9 \lesssim q < 1$. This indicates that the precise evaluation for such systems requires high computational costs.

\begin{figure*}
    \begin{tabular}{cc}
    \includegraphics[width=.48\linewidth,clip]{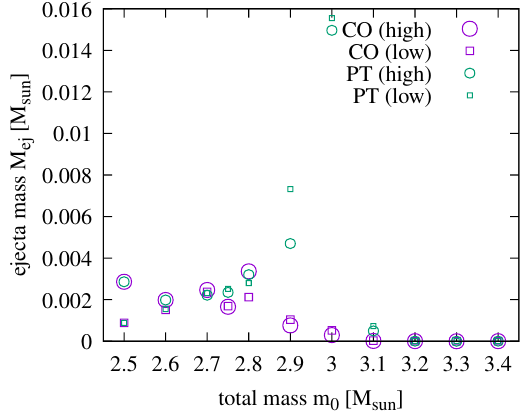} &
    \includegraphics[width=.48\linewidth,clip]{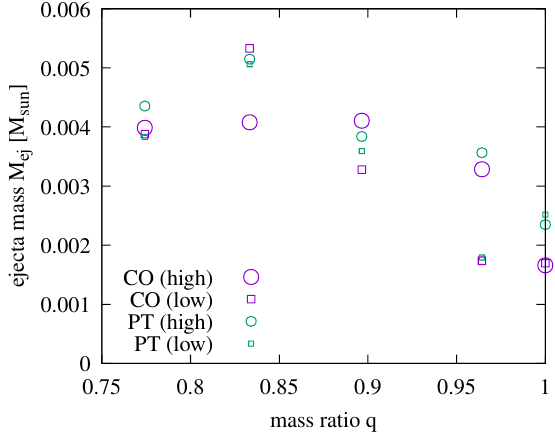}
    \end{tabular}
    \caption{Mass of the ejected material, $M_\mathrm{ej}$, as a function of the total mass of the system, $m_0$, for equal-mass models (left) and as a function of the mass ratio, $q$, for $m_0 = 2.75M_\odot$ models (right). The large purple and small green symbols represent the results for the CO and PT scenarios, respectively. Results for high ($\approx \SI{190}{\meter}$) and low ($\approx \SI{280}{\meter}$) resolutions are shown with circles and squares, respectively, to check numerical uncertainties.} \label{fig:mej}
\end{figure*}

For completeness, we present in Fig.~\ref{fig:mej} the mass of dynamical ejecta, $M_\mathrm{ej}$, as a function of the total mass, $m_0$, and the mass ratio, $q$, although $M_\mathrm{ej}$ does not reach $0.05M_\odot$ by itself only. We also include the results for the cases that a long-lived remnant is formed. Similarly to the case of $M_\mathrm{disk}$, the prompt collapse does not allow a substantial amount of material to be ejected (but see Ref.~\cite{Kiuchi:2019lls} for a massive asymmetric binary). For the cases of non-prompt collapse with $t_\mathrm{life} \lesssim \SI{30}{\ms}$, which corresponds to $2.7M_\odot \lesssim m_0 \lesssim 2.9M_\odot$ in the CO scenario, $\order{\num{e-3}}M_\odot$ is ejected with complicated dependence on $m_0$. While the dependence continues to be nonmonotonic for long-lived remnants, we find that the system with $m_0 = 3.0M_\odot$ in the PT scenario ejects $\approx 0.015M_\odot$. It appears that systems with $m_0$ marginally below the threshold of prompt collapse tend to eject a large amount of material as a result of the violent merging process. Regarding the mass ratio, dynamical mass ejection in this study tends to be efficient for an asymmetric system, while quantitative dependence of $M_\mathrm{ej}$ on $q$ seems more complicated than that of $M_\mathrm{disk}$. This complicated dependence reflects the competition between shock-driven and tidally-driven mass ejection~\cite{Hotokezaka:2012ze}.

\subsubsection{Consistency with AT\;2017gfo}

Our CO scenario is consistent with AT\;2017gfo associated with GW170817 if the system is mildly or highly asymmetric. Although the model 1.375--1.375 cannot explain AT\;2017gfo (unless nearly 100\% of the disk material at \SI{10}{\ms} after the gravitational collapse is ejected), the ejection efficiency of $\approx 30\%$ makes all the asymmetric models consistent with AT\;2017gfo. This efficiency may reasonably be realized by the effective viscosity caused by magnetohydrodynamic turbulence \cite{Kiuchi:2022nin}. In light of the distribution of $q$ for Galactic binary neutron stars \cite{Farrow:2019xnc}, our CO scenario suggests that kilonovae as bright as AT\;2017gfo will be a typical outcome of binary-neutron-star mergers with GW170817-like total masses.

Because no material disappears beyond the black-hole horizon for $m_0 = 2.75M_\odot$, the PT scenario may also be consistent with AT\;2017gfo irrespective of the mass ratio. In either scenario, however, too efficient mass ejection would produce an overluminous kilonova and is not favored unless microphysical ingredients such as the heating rate are significantly revised~\cite{Hotokezaka:2015cma,Barnes:2016umi}. In particular, if magnetar-like activity works efficiently to enhance the kinetic energy of the ejecta, the PT scenario would have produced a radio afterglow brighter than that associated with AT\;2017gfo.

\subsection{Spin of the remnant black hole} \label{sec:sim_spin}

The spin of the remnant black hole may have a sizable impact on gamma-ray bursts, which are likely to be powered by the Blandford-Znajek process \cite{Blandford:1977ds}. To motivate consideration in this direction, we list the dimensionless spin parameter $\chi$ of the remnant black hole in Table \ref{tab:model}. In this study, the spin of the black hole is estimated from the ratio of the polar to equatorial circumferential radii \cite{Smarr:1972kt}. Previous numerical experiments suggest that this approximate method is sufficient to characterize the spin magnitude and capture overall dependence on binary parameters (see, e.g., Ref.~\cite{Kyutoku:2011vz}). While measurements based on the isolated or dynamical horizon may be sophisticated alternatives \cite{Dreyer:2002mx,Schnetter:2006yt,Lovelace:2008tw}, any method has limitations to its legitimacy in dynamical spacetimes with mass accretion.

The black hole formed via prompt collapse, which occurs only for equal-mass models in our study, spins as rapidly as $\chi \approx 0.8$. This value is largely consistent with those found in previous studies \cite{Baiotti:2008ra,Kiuchi:2009jt} and significantly larger than $0.686$ found for equal-mass binary black holes \cite{Scheel:2008rj}. This is because the finite size of neutron stars terminates the inspiral phase of binary neutron stars earlier than that of binary black holes and accordingly the angular momentum emission prior to merger is less efficient. Among the cases of prompt collapse, more massive systems lead to slightly smaller values of $\chi$. This is reasonably understood to reflect the high compactness of a massive neutron star. Binaries with more massive (and more compact) neutron stars are closer to binary black holes than those with lighter ones, so that the inspiral phase is longer and the angular momentum is emitted more efficiently.

The spin parameter of a black hole formed after the collapse of a remnant massive neutron star is significantly smaller than that after the prompt collapse. This is because the angular momentum of the remnant is distributed to the surrounding material via hydrodynamic processes. Actually, Table \ref{tab:model} indicates that the value of $\chi$ is smaller for a lower-mass system with a longer remnant lifetime, which helps the angular momentum transport. Combined with the argument made above, we conjecture that the spin parameter of the remnant black hole will take a maximum value when $m_0$ is the threshold of prompt collapse. However, $M_\mathrm{disk}$ is unlikely to be substantial for the prompt collapse at least for the equal-mass system, and thus the maximum spin may not be achieved in configurations that generate realistic short-hard gamma-ray bursts.

The mass asymmetry reduces the value of $\chi$, and this is also understood as a consequence of the angular momentum transport to surrounding material. Although we investigate the effect of mass asymmetry only for the cases that massive neutron stars are formed, we speculate that the same trend holds for the case of prompt collapse based on these physical considerations above.

\section{Assessment of systematic uncertainties} \label{sec:sys}

Our merger simulations are performed incorporating only simplified physics inputs even within the scenarios advocated in Sec.~\ref{sec:eos}. In this section, we present results of supplementary simulations performed to assess systematic uncertainties. We exclusively focus on the CO scenario here. This is because the PT scenario leads to only prompt collapse with $t_\mathrm{life} \approx \SI{0}{\ms}$ or formation of a long-lived remnant with $t_\mathrm{life} > \SI{30}{\ms}$ in the models considered in this study, precluding meaningful assessment of systematic uncertainties.

\subsection{Standard and generalized piecewise polytropes} \label{sec:sys_pwp}

A potential drawback of the standard piecewise polytropic representation is that the sound speed becomes discontinuous at the boundary density, $\rho_i$. To check that this artificial discontinuity inherent in the piecewise polytropes does not bias our qualitative conclusion, we have performed preliminary simulations with a generalized piecewise polytrope \cite{OBoyle:2020qvf}. This model ensures that not only the pressure but also the sound speed is continuous at the boundary density by augmenting the pressure with additional constants $\Lambda_i$ as
\begin{equation}
    P_\mathrm{cold,g} (\rho) = K_i \rho^{\Gamma_i} + \Lambda_i \,.
\end{equation}
While the sound speed may still have cusps in this model, microscopic CO models such as quarkyonic matter are consistent with this behavior~\cite{McLerran:2018hbz}. Again, we compute the internal energy by the first law of thermodynamics assuming the continuity.

We fix the parameters of the generalized piecewise polytrope for the hadron-quark CO by mimicking the corresponding standard piecewise polytrope. Because the meaning of $\Gamma_i$ and $K_i$ is different between these two polytropic representations, adopting the same values of parameters is not appropriate. In particular, the peak (specifically, the decreasing part) of the sound speed requires a segment with $\Gamma_i < 1$ for the generalized piecewise polytrope. Because it is not uniquely determined what difference should be minimized in this comparison,\footnote{For example, we could have chosen the parameters so that the maximum mass and the radius of typical-mass neutron stars become identical.} we content ourselves with the overall resemblance between the standard and generalized piecewise polytropes; that is, the L1 norm of the pressure difference is approximately minimized in the range of $0.1n_\text{sat} < n < 10n_\text{sat}$. With keeping $K_1$, $\Gamma_1$ and $\Gamma_4$ the same as the standard piecewise polytrope for CO, other parameters are determined to be $\Gamma_2 = 3.005$, $\Gamma_3 = 0.04$, $\rho_1 = \SI{7.85e13}{\gram\per\cubic\cm}$, $\rho_2 = \SI{8.39e14}{\gram\per\cubic\cm}$, and $\rho_3 = \SI{1.206e15}{\gram\per\cubic\cm}$. We do not construct a generalized piecewise polytrope for the PT scenario.

\begin{figure}
  \includegraphics[width=.99\columnwidth,clip]{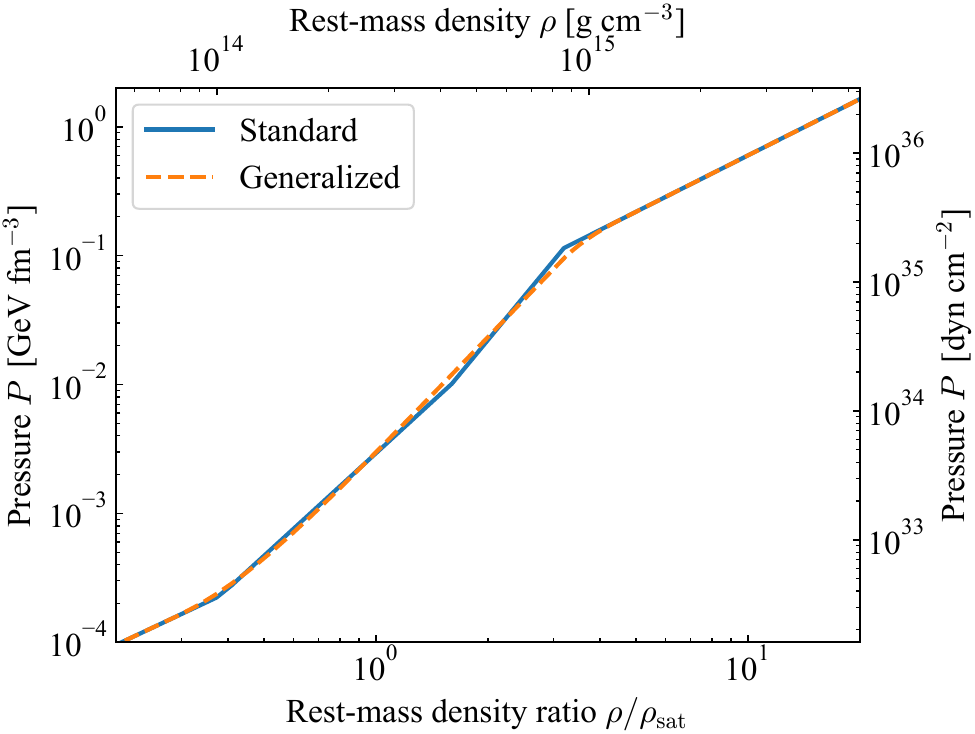}
 \caption{Comparison of our standard (blue solid) and generalized (orange dashed) piecewise polytropic representations for the EoS in the CO scenario. The vertical axis for the pressure $P$ shows values in the high-energy friendly unit, [\si{\giga\eV\per\cubic\femto\meter}], and in the astrophysics friendly unit, [\si{dyn.\cm^{-2}}]. The horizontal axis has also values in the unit of the rest-mass density at nuclear saturation, $\rho/\rho_\mathrm{sat}$, and in the physical unit, [\si{\gram\per\cubic\cm}].}
 \label{fig:pwpcompare_eos}
\end{figure}

\begin{figure}
  \includegraphics[width=.92\columnwidth,clip]{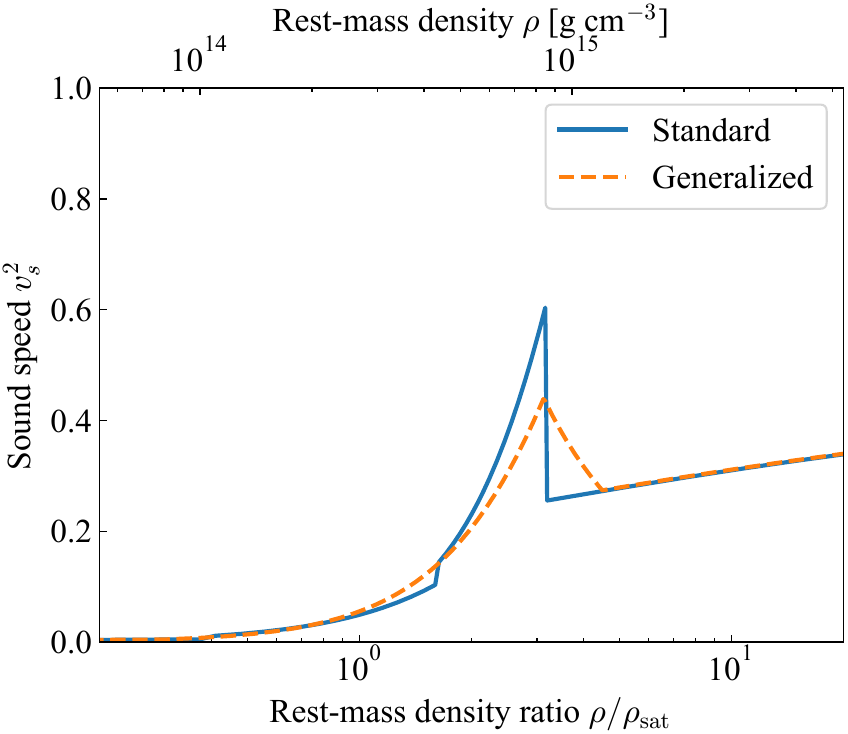}
 \caption{Comparison of our standard (blue solid) and generalized (orange dashed) piecewise polytropic representations for the sound speed. The latter exhibits cusps but no discontinuity.} \label{fig:pwpcompare_vs}
\end{figure}

Figure~\ref{fig:pwpcompare_eos} compares our standard and generalized piecewise polytropic representations for the EoS in the CO scenario. While the two lines of the EoSs closely follow each other in the entire range of $\rho$ as seen in Fig.~\ref{fig:pwpcompare_eos}, the sound speed in Fig.~\ref{fig:pwpcompare_vs} is continuous for the generalized piecewise polytrope everywhere. Nevertheless, a prominent peak still remains before approaching the pQCD branch. It should be noted that the maximum mass $M_\text{max}$ is reduced from $M_\text{max}\simeq 1.98M_\odot$ (standard) to $M_\text{max} \simeq 1.93M_\odot$ (generalized). This will later turn out to affect the lifetime of the merger remnant and observable signatures at not the qualitative but only the quantitative level.

\begin{figure*}
 \begin{tabular}{cc}
  \includegraphics[width=.48\linewidth,clip]{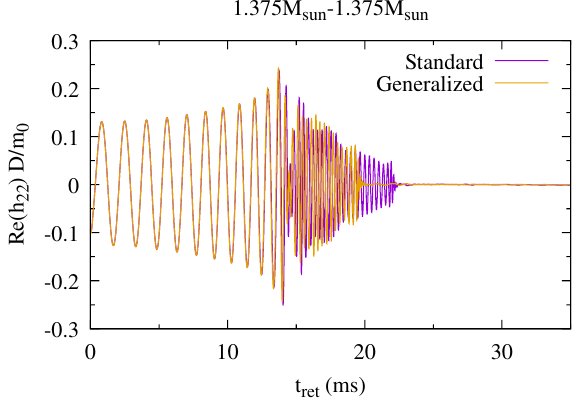} &
  \includegraphics[width=.48\linewidth,clip]{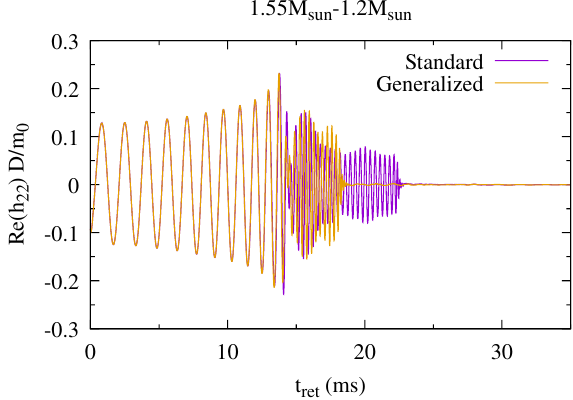}
 \end{tabular}
 \caption{Gravitational waveform obtained with standard (purple) and generalized (orange) piecewise polytropes for the models 1.375--1.375 (left) and 1.55--1.2 (right). All the results are derived in high-resolution simulations for the CO scenario.} \label{fig:gwpwp}
\end{figure*}

Figure~\ref{fig:gwpwp} compares the gravitational waveforms for the models 1.375--1.375 and 1.55--1.2 derived with standard and generalized piecewise polytropes. Except that the collapse sets in earlier by $\approx \SI{5}{\ms}$ for our generalized piecewise polytrope, no qualitative difference is found between two polytropic representations. The difference in the lifetime of the remnant may safely be ascribed to the difference in the maximum mass described above. Hence, it is natural that the collapse sets in earlier for our generalized piecewise polytrope. This comparison suggests that the standard piecewise polytropes can be used acceptably for the purpose of investigating the CO scenario in binary-neutron-star mergers despite associated discontinuities in the sound speed.

\begin{figure*}
 \begin{tabular}{cc}
  \includegraphics[width=.48\linewidth,clip]{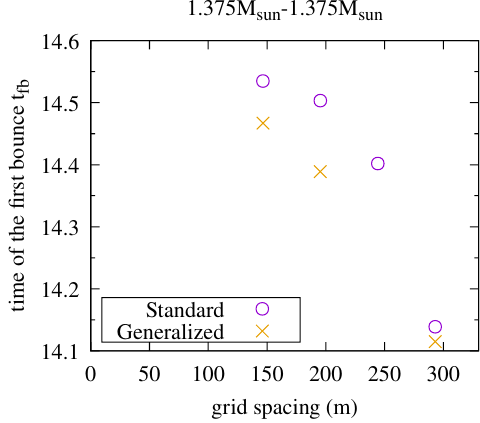} &
  \includegraphics[width=.48\linewidth,clip]{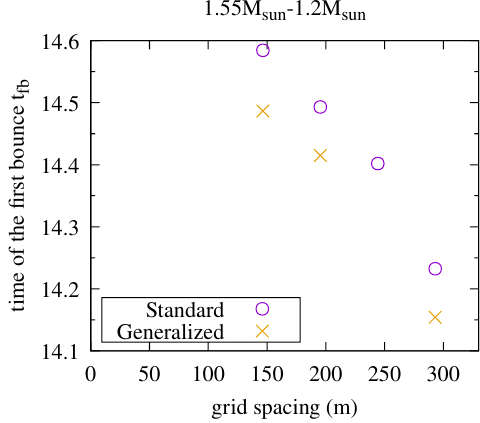}
 \end{tabular}
 \caption{Time of the first bounce, $t_\mathrm{fb}$, as a function of the grid spacing obtained with standard (purple circle) and generalized (orange cross) piecewise polytropes for the models 1.375--1.375 (left) and 1.55--1.2 (right).}
\label{fig:conv_bounce}
\end{figure*}

We comment that numerical convergence appears better-behaved for the generalized piecewise polytrope due presumably to the absence of sound-speed discontinuity. Figure~\ref{fig:conv_bounce} shows the time of the first bounce, $t_\mathrm{fb}$, for two piecewise polytropic representations as a function of grid spacings. Because the shock interaction prohibits clean convergence after the collision, we only focus on the inspiral phase up to the first bounce. With the generalized piecewise polytrope, the orders of convergence estimated from ultrahigh, high, and low resolutions\footnote{We have not performed middle-resolution simulations with the generalized piecewise polytrope.} are $2.56$ for the model 1.375--1.375 and $2.67$ for the model 1.55--1.2. A noteworthy feature is the closeness of the convergence order between two binary models. With the standard piecewise polytrope, the orders of convergence are as distinct as $5.78$ for 1.375--1.375 and $2.00$ for 1.55--1.2. Moreover, in light of various routines with different accuracy implemented in our numerical code, {\small SACRA}, the beyond-fourth-order convergence order, e.g., $5.78$, must be accidental (see, e.g., Ref.~\cite{Kyutoku:2014yba}). This irregular convergence property for standard piecewise polytropes requires a large number of simulations with different grid resolutions to take the continuum limit \cite{Kiuchi:2017pte,Kiuchi:2019kzt}. The finding here encourages us to employ generalized piecewise polytropes for deriving accurate gravitational waveforms at least for the inspiral phase. Having said that, because we have simulated only two binary models for comparing numerical convergence, it is necessary to investigate a wide range of binary parameters and underlying EoS to draw a definitive conclusion (see, e.g., Refs.~\cite{Foucart:2019yzo,Raithel:2022san,Knight:2023kqw} for related work).

\subsection{Finite-temperature effect} \label{sec:sys_therm}

The value of $\Gamma_\mathrm{th}$ for the finite-$T$ part of the EoS may generally depend on the density~\cite{Constantinou:2015mna,Carbone:2019pkr,Fujimoto:2021dvn}. However, the current prescription requires it to be constant in each simulation. To assess the uncertainty associated with this approximation, we also perform simulations with $\Gamma_\mathrm{th}=1.5$, which is likely to bracket realistic ranges of thermal effects in combination with $\Gamma_\mathrm{th}=1.75$ adopted so far \cite{Bauswein:2010dn,Figura:2020fkj}. In the present work, we do not aim at systematic investigation of the finite-$T$ contribution to the hadron-quark transition (see Refs.~\cite{Most:2018eaw,Prakash:2021wpz,Blacker:2023afl} for studies in this direction). It would be an interesting and important future extension to explore microscopic dynamics including PT in the finite-$T$ regions.
\begin{figure}
    \begin{tabular}{c}
    \includegraphics[width=.95\linewidth,clip]{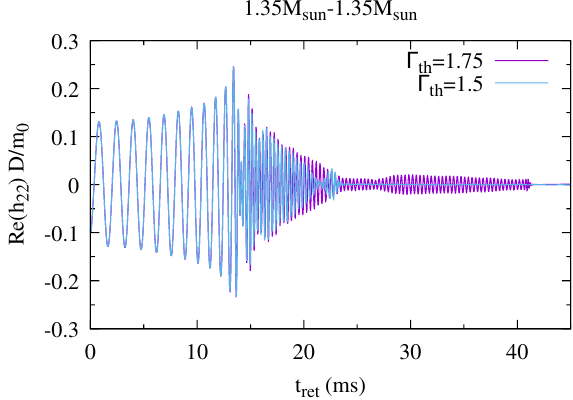} \\
    \includegraphics[width=.95\linewidth,clip]{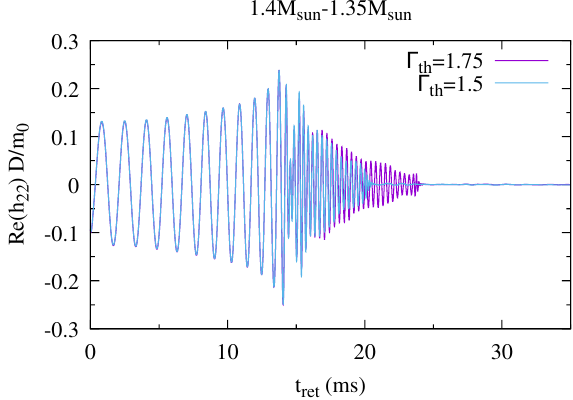} \\
    \includegraphics[width=.95\linewidth,clip]{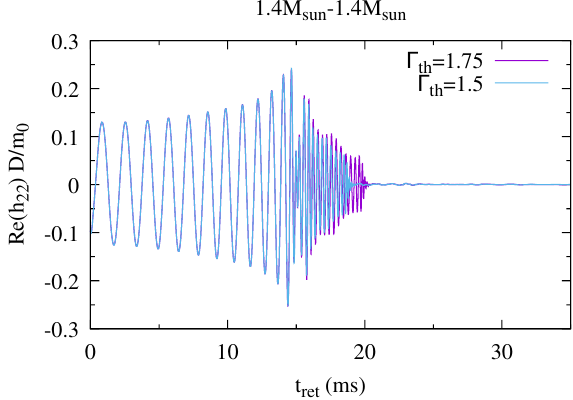}
    \end{tabular}
    \caption{Gravitational waveform obtained with $\Gamma_\mathrm{th} = 1.75$ (purple) and $1.5$ (cyan) for the models 1.35--1.35 (top), 1.4--1.35 (middle), and 1.4--1.4 (bottom). All the results are derived in high-resolution simulations performed in the CO scenario.} \label{fig:gwgamth}
\end{figure}

Figure~\ref{fig:gwgamth} shows the gravitational waveforms for the models 1.35--1.35 (top), 1.4--1.35 (middle), and 1.4--1.4 (bottom) derived with two choices of $\Gamma_\mathrm{th}$, i.e., $\Gamma_\mathrm{th}=1.75$ presented in Sec.~\ref{sec:sim} and $\Gamma_\mathrm{th}=1.5$ for comparison. In all the models, a smaller value of $\Gamma_\mathrm{th} = 1.5$ makes the lifetime, $t_\mathrm{life}$, of the remnant shorter. This is because the weak finite-temperature effect characterized by a small value of $\Gamma_\mathrm{th}$ reduces the thermal pressure in the postmerger system, accelerating the gravitational collapse.

Quantitatively, the longer $t_\mathrm{life}$ is, the more $t_\mathrm{life}$ varies. For a relatively short-lived remnant with $t_\mathrm{life} \lesssim \SI{10}{\ms}$ in the models 1.4--1.35 and 1.4--1.4, the variation is $\lesssim \SI{5}{\ms}$ as seen in the middle and bottom panels of Fig.~\ref{fig:gwgamth}. This amount of variation is comparable to truncation errors, particularly for asymmetric systems. For a relatively long-lived remnant with $t_\mathrm{life} \approx \SI{30}{\ms}$ in the model 1.35--1.35, the variation is also as long as $\approx \SI{15}{\ms}$. This indicates that a longer-lived remnant will be affected more significantly by the strength of finite-temperature effects. We also comment that a larger value of $\Gamma_\mathrm{th}$, e.g., probably unrealistic choice of $\Gamma_\mathrm{th}=2$, will lead to a significantly long-lived remnant (see, e.g., Ref.~\cite{Huang:2022mqp}), although neutrino emission will eventually deprive the remnant of the thermal support and induce gravitational collapse.

These results may be summarized to the idea that the strength of finite-temperature effects will modify the informative mass range that allows us to distinguish two transition scenarios via black-hole formation during intense gravitational-wave emission. Still, because the variation is small for the models 1.4--1.35 and 1.4--1.4, we may robustly distinguish these scenarios with binary neutron stars with $m_0$ in an astrophysically promising mass range discussed in Sec.~\ref{sec:eos}. To delineate transition behavior precisely, it is desired to elaborate our understanding of finite-temperature effects further including their influence on hadron-quark transition itself~\cite{Most:2018eaw,Prakash:2021wpz,Blacker:2023afl}, as the temperature dependence of its phase boundary makes it occur more easily at finite temperatures and softens the finite-temperature EoS.
 
\section{Summary and future prospect} \label{sec:summary}

We explored observational signatures of hadron-quark transition in binary-neutron-star mergers by performing numerical-relativity simulations. Extending our previous \textit{Letter} \cite{Fujimoto:2022xhv}, various binary configurations are studied systematically under the assumption of either hadron-quark crossover (CO) or strong first-order phase transition (PT). By construction, the EoSs are identical in the density range relevant to premerger neutron stars, so that the inspiral waveforms are identical. Accordingly, we focus mainly on the postmerger phase.

We strengthened our previous finding that softening toward the pQCD branch in the CO scenario induces earlier gravitational collapse of the postmerger remnant than in the PT scenario. The short lifetime of the remnant for CO is reflected in gravitational-wave signals. In particular, they suddenly shut off if the gravitational collapse occurs while the emission is strong. The lifetime depends primarily on the total mass of the system, $m_0$, and only weakly on the mass ratio. The difference between the two scenarios is remarkable in the mass range $m_0 \approx 2.7$--$2.9M_\odot$, which nicely overlaps with the expected distribution of astrophysical binary neutron stars. If our CO scenario is correct, GW170817 should have formed a black hole after $\approx \SI{10}{\ms}$ after merger. Future detectors may enable us to clarify the physical transition scenario \cite{Harada:2023eyg}.

Multimessenger observations will also be useful for delineating the hadron-quark transition. The mass of the accretion disk surrounding the formed black hole is sensitive to the lifetime of the remnant massive neutron star. Although our simulations are not adequate to investigate postmerger mass ejection, this sensitivity would also be reflected in the amount of disk outflow and the luminosity of associated electromagnetic emission. Our CO scenario is consistent with the brightness of AT\;2017gfo if the binary was even mildly asymmetric. For the PT scenario, long-lived remnants are typical outcomes so that the emission would tend to be bright. As usual, dynamical ejecta are not sufficient to explain AT\;2017gfo in any model. The spin of the remnant black hole, which is critical for the jet launch, also depends on how the gravitational collapse occurs. From our systematic investigation with various mass configurations, we conjecture that the dimensionless spin takes the maximal value of $\gtrsim 0.8$ at the threshold mass of the prompt collapse for a given EoS.

We also performed auxiliary simulations to check the robustness of our main conclusion with respect to details of numerical implementations. The generalized piecewise polytrope is adopted to confirm that the dynamics and gravitational waves are not biased artificially by the discontinuity of the sound speed inherent to the standard piecewise polytrope. Strength of the thermal effect is varied within an realistic range to confirm that its influence on the lifetime is minor for the short-lived remnants relevant to the scenario distinction. Although we did not investigate the magnetic field and the neutrino transport expecting that they play an important role only at late times of $\gtrsim \SI{100}{\ms}$ after merger, they will ultimately affect the lifetime of even short-lived remnants to some extent.

Our quantitative conclusion depends on the physical assumption on the hadron-quark transition scenario. In particular, the assumed maximum mass of $\approx 2M_\odot$ for the CO scenario is essential for the specific range of the total mass covering GW170817, $m_0 \approx 2.7$--$2.9M_\odot$, to be promising for the scenario distinction. Although the uncertainty is large, some pulsar observations suggest that the maximum mass of the neutron star could significantly exceed $2M_\odot$, as well as some model-dependent analysis of multimessenger astronomical observations. If the maximum mass is high in reality, a GW170817-like system may also form a long-lived remnant, and accordingly postmerger gravitational waves will not be helpful for the scenario distinction. Instead, more massive systems like GW190425 will result in a short-lived remnant that allow us to identify black-hole formation and will play a more important role for this purpose. Detailed investigations of the relation between the maximum mass and the useful mass range for the scenario distinction will be left for the future study. We recall that the EoS at intermediate density will be determined to high precision via future observations of inspiral gravitational-wave signals. We may safely expect that the uncertainty in theoretical scenarios will be narrowed down before the postmerger gravitational-wave signal becomes available to us.

\begin{acknowledgments}
 The authors thank Len Brandes and Toru Kojo for discussions. This work was supported by JSPS KAKENHI Grant-in-Aid for Scientific Research No.~JP19K14720 (KK), No.~JP20H00158 (KH), No.~JP20H05639 (KH), No.~JP22H01216 (KF), No.~JP22K03617 (KK), No.~JP22H05118 (KF), No.~JP23H01169 (KH), No.~JP23H04900 (KH), JST FOREST Program No.~JPMJFR2136 (KH), and the work of YF was supported by the Institute for Nuclear Theory's U.S. DOE Grant No. DE-FG02-00ER41132. YF, KF, and KK acknowledge the Yukawa Institute for Theoretical Physics at Kyoto University and RIKEN iTHEMS, where part of this work was completed during the International Molecule-type Workshop ``Condensed Matter Physics of QCD 2024'' (YITP-T-23-05).
\end{acknowledgments}

%

\end{document}